\documentclass[12pt,letterpaper]{article}

\usepackage{amsmath,amssymb,amsthm}
\usepackage{mathtools}
\usepackage{geometry}
\usepackage{graphicx}
\usepackage{hyperref}
\usepackage{booktabs}
\usepackage{enumitem}
\usepackage{array}
\usepackage{mathrsfs} 
\usepackage[table]{xcolor}
\usepackage{pifont} 
\usepackage{tikz}
\usepackage{tikz-cd}
\usetikzlibrary{decorations.pathreplacing}
\usepackage{pgfplots}\pgfplotsset{compat=1.16}\usepgfplotslibrary{groupplots}

\hypersetup{colorlinks=true,linkcolor=blue,citecolor=blue,urlcolor=blue}

\newcommand{\comment}[1]{}
\newcommand{\CC}{\mathbb{C}}

\newcommand{\ZZ}{\mathbb{Z}}
\newcommand{\HH}{\mathcal{H}}
\newcommand{\ket}[1]{|#1\rangle}
\newcommand{\bra}[1]{\langle#1|} 
\newcommand{\im}{\operatorname{im}}
\newcommand{\rk}{\operatorname{rk}}
\newcommand{\coker}{\operatorname{coker}}

\newcommand{\Tr}{\operatorname{Tr}}

\newcommand{\cH}{\mathcal{H}}
\newcommand{\cM}{\mathcal{M}}

\newcommand{\Dtil}{\widetilde{D}}

\newcommand{\W}{\mathcal{W}}

\newcommand{\cF}{\mathcal{F}}

\theoremstyle{plain}
\newtheorem{theorem}{Theorem}[section]

\newtheorem{conjecture}[theorem]{Conjecture}

\theoremstyle{definition}

\theoremstyle{remark}

\newcommand{\Bps}{\mathcal{B}}
\newcommand{\Npsi}{N_\psi}
\newcommand{\Nchi}{N_\chi}

\newcommand{\usefig}[2][\textwidth]{%
  \IfFileExists{#2.pdf}%
    {\includegraphics[width=#1]{#2.pdf}}%
    {\resizebox{#1}{!}{\input{#2.tex}}}%
}

\ifdefined\pdfshellescape
  \ifnum\pdfshellescape=1
    \immediate\write18{bash build_figures.sh}%
  \else
  \fi
\else
\fi

\begin{document}

\begin{titlepage}
    \begin{flushright}
        QMUL-PH-26-29\
    \end{flushright}
    \null
    \vspace*{\fill}

\begin{center}

\centerline{\Large \bf Fortuity and fragility in supersymmetric SYK} 

\bigskip
\bigskip

{James Chryssanthacopoulos, David Vegh}

\bigskip

{\it Centre for Theoretical Physics and Astronomy, Queen Mary University of London,\\
Mile End Road, London E1 4NS, United Kingdom}

\medskip

email: {\tt j.chryssanthacopoulos@qmul.ac.uk, d.vegh@qmul.ac.uk}

\bigskip

{\it \today}

\bigskip

\begin{abstract} 
\noindent
In the $\mathcal N=2$ supersymmetric SYK model with $N$ fermions, every BPS state is \emph{fortuitous}: at fixed charge, it exists only over a finite range of~$N$. Allowing the charge to increase, we show that BPS classes may instead be uplifted along lattice walks inside the fortuity window, potentially to arbitrarily large~$N$. However, there is no canonical uplift. We therefore introduce a decoder $D$ whose spectrum quantifies exact uplift and its metric cost. We call the failure or increasing cost of uplifting harmonic representatives \emph{metric fragility}. 
Chen's single-matrix model and the protected tower of the two-flavor SYK model realize bare and genuinely dressed mechanisms of perfect uplift, respectively. In the generic one-flavor model, exact uplift eventually fails. The Lin--Maldacena--Rozenberg--Shan-type operator $D^\dagger D$ exhibits level statistics consistent with the Gaussian unitary ensemble. Its eigenvalues quantify the metric continuity of BPS states across system size, while their correlations probe chaos within the BPS sector. Metric fragility and BPS chaos are thus encoded in complementary observables of the same operator, distinguishing chaotic BPS sectors from exactly solvable towers.

\end{abstract}

\end{center}
  \vspace*{\fill}
\end{titlepage}

\tableofcontents

\clearpage
\section{Introduction}
\label{sec:intro}

A striking lesson of recent work on supersymmetric theories is that a
large class of BPS microstates conjectured to be holographically dual to black holes exists only at finite~$N$. In the classification of Chang and Lin~\cite{ChangLin}, rooted in the
supercharge cohomology approach to $\mathcal{N}=4$ super
Yang--Mills~\cite{ChangYin}, a BPS state of a family of theories labeled by
the rank parameter $N$ is \emph{monotonous} if it descends from a
supersymmetric state of the $N\to\infty$ theory. Instead, a state is \emph{fortuitous} if its
supersymmetry relies on finite-$N$ relations, so that the state exists only
within a finite range of consecutive ranks. Monotonous states are naturally
associated with smooth horizonless configurations such as graviton gases,
while the exponentially numerous fortuitous states are the candidate
microstates of supersymmetric black holes~\cite{ChangLin,CCSY}. The distinction has by now been studied in the $\mathcal N=2$ SYK model~\cite{CCSY}, in a solvable single-matrix model~\cite{Chen}, in the D1--D5 system~\cite{ChangLinZhang,HughesShigemori,Hughes:2026qqn, Hughes:2026naj, Giusto:2026rpl}, in higher-spin and bosonic vector models~\cite{Kim:2025vup, deMelloKoch:2025cec}, in quark models through BRST cohomology~\cite{Fliss:2026rad}, and in ABJM theory~\cite{ABJMfortuity}.  

Why track individual microstates across system sizes? The question of whether a given black hole microstate retains any identity when the theory itself is enlarged is a sharp version of the question of what black hole microstates are.  
If fortuitous states simply flicker in and out of existence as $N$ varies, 
microstate identity across theories is meaningless. If instead each state admits a controlled uplift, the family of theories carries a coherent tower of microstates, and one can ask
quantitatively how much of a state survives the uplift to large $N$. Related approaches to following states as the rank varies were developed in~\cite{Budzik:2023vtr}.

In this paper, we construct this uplift for the $\mathcal{N}=2$ SYK model~\cite{FGMS,CCSY}, a supersymmetric variant of the SYK model~\cite{SachdevYe,Kitaev,MaldacenaStanford}: $N$ complex fermions with a random $q$-body supercharge $Q_N$, whose BPS states are the harmonic representatives of the $Q_N$-cohomology. For generic couplings the cohomology is confined to the \emph{fortuity window} $|2p-N|\leq q$, where $p$ is the fermion number, or $R$-charge. Every BPS state is fortuitous: held at fixed~$p$, it falls out of the window as $N$ grows and is destroyed~\cite{CCSY}. This $R$-charge concentration has a low-energy counterpart in the $\mathcal N=2$ Schwarzian~\cite{Stanford:2017thb,CCSY}. Our starting point is the observation that a state may nevertheless survive indefinitely if its fermion number is allowed to drift upward with~$N$.

In~\cite{CCSY}, the BPS spectra at $N$ and $N{+}1$ were related by a long exact sequence in cohomology. Iterating it mode by mode, we label
uplifts of BPS classes by binary strings (lattice \emph{walks}) in which the
$k$-th bit records whether the fermion number is unchanged (a \emph{down
step}) or it is increased by one (an \emph{up step}).
We show that BPS classes may admit unobstructed cohomological
extensions to arbitrarily large~$N$ along a walk inside the fortuity window. However, uplifting is intrinsically ambiguous.
First, the long exact sequence determines the uplift only as an element of a quotient of the rank-$N{+}1$ cohomology. Choosing a representative of the resulting coset selects a particular cohomology class, and different representatives may have different uplift properties at subsequent ranks. Moreover, there is no canonical, purely algebraic map from harmonic
states at $N$ to harmonic states at $N{+}1$. 

In this paper, we attempt to quantify uplifting using the inner product. For a
set $M$ of added modes and a designated occupation pattern $T\subseteq M$, we
define the \emph{decoder} $D_T:\mathcal B_{N'}^P\to\HH_N^p$, which maps the BPS states $\mathcal B_{N'}^P$ in a given charge sector $P$ and system size $N'$ into the Hilbert space $\HH_N^p$ of the smaller theory ($N<N'$ and $p$ is the fixed charge sector).\footnote{
Earlier studies of quantum error correction in SYK addressed reconstruction after the erasure of subsets of fermions and interpreted low-energy subspaces as approximate quantum codes \cite{Chandrasekaran:2022qmq,Bentsen:2023xlu}. Our decoder addresses a different question: whether BPS states can be continued between neighboring system sizes after resolving the occupation of the removed modes.}  The decoder's image is the set of old states that occur
exactly as the $T$-component of a genuine BPS state upstairs. We define the
\emph{uplift fidelity} $\cF_T$ as the fraction of the state's squared norm contained in the decoder image.   Sometimes, a state $\alpha$ can be uplifted by simply taking the product state $\alpha\otimes\ket{T}$. However, the resulting state may not be harmonic and corrections must be added to restore harmonicity. The \emph{dressing cost}  $\kappa_T$ captures the fractional norm overhead of the correction, which vanishes exactly when $\alpha\otimes\ket{T}$ is already harmonic.

The notions of uplift fidelity and dressing cost help distinguish between fortuitous and monotonous behavior in three models:
\begin{enumerate}[label=(\roman*),itemsep=2pt]
\item In the single-matrix model of Chen~\cite{Chen}, one can uplift every BPS state of
rank~$N$ to a BPS state of rank $N{+}1$ with no correction at all. The
diagnostics take rigid values, $\cF_T=1$ and $\kappa_T=0$. The model is a 
rigid ``skeleton'' of fortuity.

\item In the symmetrized two-flavor model of~\cite{CCSY, Heydeman:2022lse}, the monotonous states $V^n\ket{\Omega_N}$ built
from the bilinear $V=\sum_i\psi_i\chi_i$ lift exactly, $\cF_T=1$,
with a nonvanishing correction fixed by the symmetry $[Q,V]=0$ and
a dressing cost bounded by $n/(N-n+1)$, which vanishes at fixed charge
as $N$ grows.

\item In the generic $\mathcal N=2$ SYK model, exact uplift of fortuitous states eventually fails outright, $\cF_T<1$. We call this \emph{metric fragility} because it is invisible to cohomology alone: it depends on the inner-product geometry of the Hilbert space, which defines harmonic representatives, orthogonal projections, and minimum-norm uplifts.
\end{enumerate}

The operator constructed from the decoder $D_T^\dagger D_T$ maps the BPS space of the enlarged theory at $N+1$ to itself. This is automatically an operator of the type studied by~\cite{LMRS,ChenLinShenker,Miyahara:2026iso}, which was claimed to probe the chaos of the BPS sector. The properties of $D_T^\dagger D_T$ can also be studied in the spirit of this program, comparing to predictions from random matrix theory. This provides a bridge connecting the concepts of fortuity, metric fragility, and BPS chaos.

The organization of the paper is as follows. Section~\ref{sec:model} reviews the $\mathcal{N}=2$ SYK model, its Witten
index, and the fortuity window. Section~\ref{sec:fortuity} reviews the
one-fermion long exact sequence of~\cite{CCSY} in a self-contained block
formalism. Section~\ref{sec:walks} constructs the walk labeling, shows
that an uplift is always possible, and isolates the 
ambiguities of uplifting. Section~\ref{sec:rayleigh} defines the
decoder and its diagnostics. Section~\ref{sec:solvable} discusses the fidelity and dressing cost in the two-flavor and single-matrix models. Section~\ref{sec:generic} studies the generic  SYK
model. Section~\ref{sec:discussion} discusses interpretation and open problems.
Appendix~\ref{app:walks} contains the walk counting of the Hodge sectors.

\section{$\mathcal{N}=2$ supersymmetric SYK model}
\label{sec:model}

Consider $N$ complex fermions $\psi_1,\dotsc,\psi_N$, satisfying the
canonical anticommutation relations
\begin{equation}\label{eq:CAR}
  \{\psi_i,\psi_j^\dagger\}=\delta_{ij},\qquad
  \{\psi_i,\psi_j\}=0,\qquad
  \{\psi_i^\dagger,\psi_j^\dagger\}=0.
\end{equation}
The Hilbert space is the Fock space built from a vacuum state
$\ket{0}$ annihilated by all $\psi_i^\dagger$:
\begin{equation}\label{eq:Hilbert}
  \HH_N=\bigoplus_{p=0}^{N}\HH_N^p,\qquad
  \HH_N^p \cong  \Lambda^p(\CC^N),\qquad
  \dim\HH_N^p=\binom{N}{p}.
\end{equation}
Here $p$ is the \emph{fermion number}, the eigenvalue of the
number operator $\hat{F}=\sum_{i=1}^N\psi_i\psi_i^\dagger$. The
total Hilbert space has dimension $\dim\HH_N=2^N$.
An arbitrary $p$-fermion state $|\alpha_p\rangle$ can be identified with a $p$-form
\(\alpha \in \Lambda^p(\mathbb{C}^N)\):
\begin{equation}
|\alpha_p\rangle = \sum_{1 \le i_1 < \cdots < i_p \le N}
\alpha_{i_1\cdots i_p} \, \psi_{i_1} \cdots \psi_{i_p} \, |0\rangle
   \, ,
\end{equation}
with fully antisymmetric coefficients \(\alpha_{i_1\cdots i_p}\).

For general odd $q$, the supercharge is a $q$-body operator
\begin{equation}\label{eq:Q-general}
  Q_N = \sum_{1\le i_1<\cdots<i_q\le N}C_{i_1\cdots i_q}\,
  \psi_{i_1}\cdots\psi_{i_q},
\end{equation}
where the coupling constants $C_{i_1\cdots i_q}$ are drawn
independently from a complex Gaussian distribution with zero mean
and variance $\langle|C_{i_1\cdots i_q}|^2\rangle
=J(q-1)!/N^{q-1}$, matching the conventions
of~\cite{FGMS,CCSY}. The scale $J$ sets the overall normalization of the
Hamiltonian and plays no role in what follows: all quantities we
introduce are invariant under $Q_N\to\lambda Q_N$. 
Defining $Q^p_N \equiv Q_N\big|_{\HH_N^p}$, we have $Q^p_N\colon\HH_N^p\to\HH_N^{p+q}$.
For any choice of couplings, the supercharge
satisfies $Q_N^2=0$, i.e.\ $Q^{p+q}_N \circ Q^p_N = 0$ for all~$p$.
The Hamiltonian is positive semi-definite
\begin{equation}\label{eq:Hamiltonian}
  H=\{Q_N,Q_N^\dagger\} \succeq 0 \, .
\end{equation} 
This defines the $\mathcal{N}=2$ supersymmetry algebra.
A state $\ket{\Psi}\in\HH_N^p$ is \emph{BPS} if $H\ket{\Psi}=0$,
which by the positivity of~\eqref{eq:Hamiltonian} is equivalent to
$Q_N\ket{\Psi}=0$ and $Q_N^\dagger\ket{\Psi}=0$.
The space of BPS states at fermion number~$p$ is canonically
isomorphic to the $Q_N$-cohomology:
\begin{equation}\label{eq:Hodge}
  \mathcal B_N^p \cong
  H^p(Q_N)\coloneqq\frac{\ker Q^p_N}
  {\im Q^{p-q}_N}.
\end{equation}
The standard Hodge decomposition of the Hilbert space is given by
\begin{equation}\label{eq:hodge-decomp}
\HH_N^p=\mathcal B_N^p
\;\oplus\;\im\, Q^{p-q}_N
\;\oplus\;\im\, (Q_N^\dagger)^{p+q}.
\end{equation}

\begin{figure}[tb]
\begin{center}
\includegraphics[width=12cm]{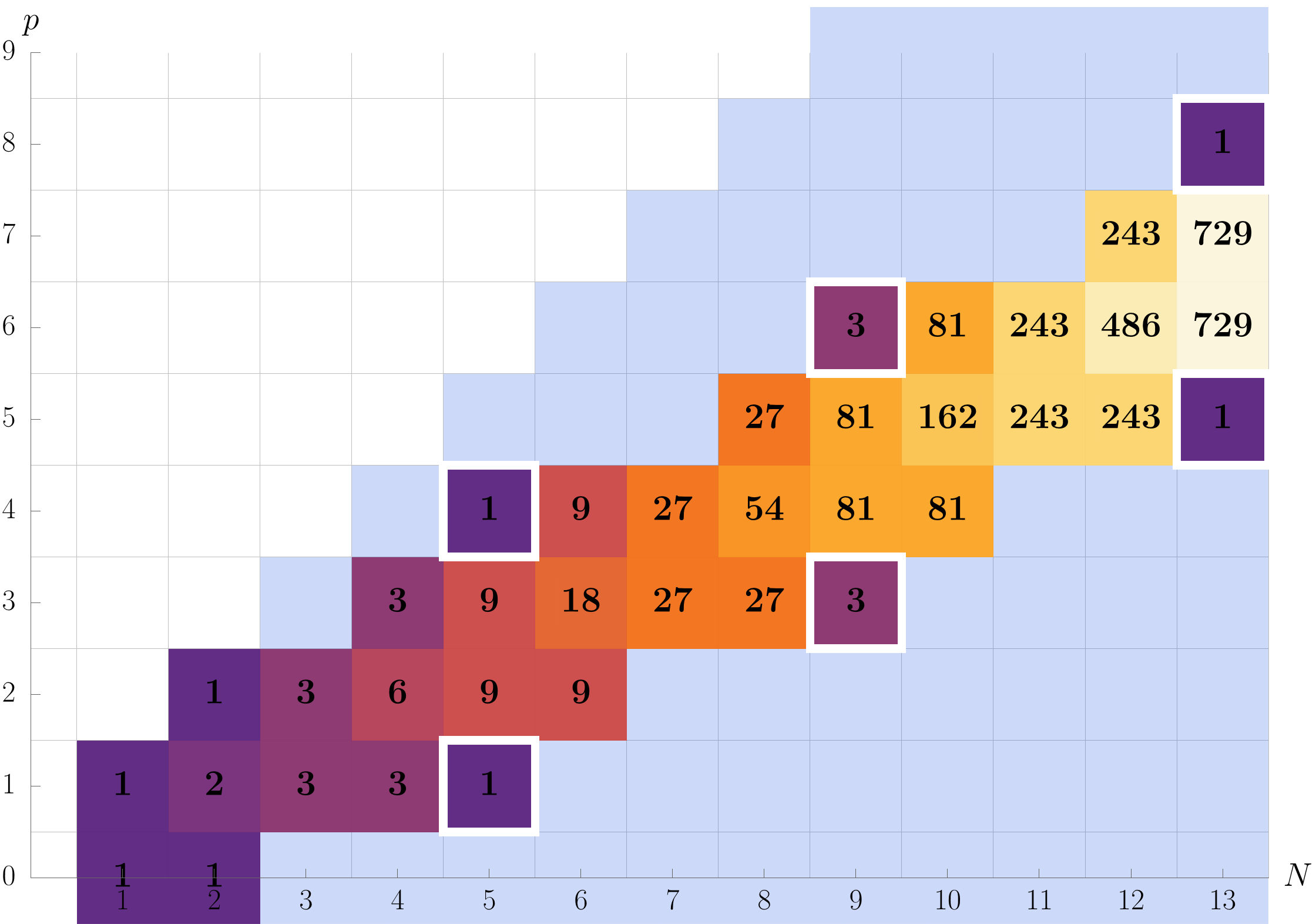}
\caption{\label{fig:bpsdata}
Dimensions of $H^p(Q_N)$ in the $\mathcal N=2$ supersymmetric SYK model with $q=3$ and generic couplings. The values were obtained by exact diagonalization of one generic coupling realization at each~$N$ and are constant across generic realizations. Exceptional states, which are invisible to the Witten index, are indicated by white squares.}
\end{center}
\end{figure}

\subsection{The Witten index}

Let $\omega=e^{2\pi i/q}$. The $\ZZ_q$ transformation
$g\colon\psi_i\mapsto\omega\psi_i$ commutes with $Q_N$ (since
the supercharge contains $q$ factors of~$\psi$). The cochain complex $(\HH_N,Q_N)$ therefore decomposes into $q$ independent
sub-complexes labeled by $r=p\bmod q$:
\begin{equation}\label{eq:Zq}
  C_r\colon\;\;
  \HH_N^{r}\xrightarrow{Q_N}\HH_N^{r+q}
  \xrightarrow{Q_N}\HH_N^{r+2q}\to\cdots
\end{equation}
The $n$-th term of this complex is
$\HH_N^{qn+r}=\Lambda^{qn+r}(\CC^N)$, of dimension
$\binom{N}{qn+r}$.

The ordinary Witten index vanishes identically:
$\operatorname{Tr}(-1)^F=\sum_p(-1)^p\binom{N}{p}=0$.
The contribution to the Witten index from the sector~$r$
is evaluated via the standard root-of-unity filter. Up to an overall sign,
\begin{equation}\label{eq:index-sector}
  \chi(r)
  =\sum_{\substack{p\ge0\\p\equiv r\!\!\!\pmod{q}}}
  (-1)^p\binom{N}{p}
  =\sum_{p=0}^N   (-1)^p\binom{N}{p}
  \left[  {1\over q} \sum_{j=0}^{q-1}\omega^{j(p-r)} \right]
  \, .
\end{equation}
We get
\begin{equation}\label{eq:root-filter}
  \chi(r)  =\frac{1}{q}\sum_{j=0}^{q-1}\omega^{-jr}
  \underbrace{\sum_{p=0}^{N}(-\omega^j)^p\binom{N}{p}}_{W_j}
  =\frac{1}{q}\sum_{j=0}^{q-1}\omega^{-jr}\,(1-\omega^j)^N,
\end{equation}
where $W_j=\operatorname{Tr}[(-1)^F g^j]$ is
the $\ZZ_q$-twisted Witten index.
The $j=0$ term vanishes, and the remaining $q-1$ terms can be
grouped into complex conjugate pairs.
For $q=3$, the sum evaluates to
\begin{equation}\label{eq:index-explicit-q3}
  \chi(r)\Big|_{q=3}
  =\frac{2}{3}\cdot 3^{N/2}\,(-1)^{r}\cos\Bigl(
  \frac{\pi(N-2r)}{6}\Bigr).
\end{equation}
The nonvanishing sector indices grow as $3^{N/2}$,
providing an exponentially large lower bound on the BPS degeneracy, with entropy
$S_0/N=\frac{1}{2}\ln 3$ \cite{FGMS}.

\subsection{Fortuity window and exceptional modes}
\label{sec:exceptional}

For generic couplings $C_{i_1\cdots i_q}$ the cohomology $H^p(Q_N)$ is expected to vanish outside the \emph{fortuity window}~\cite{CCSY},
\begin{equation}
    \label{eq:window}
    \frac{N-q}{2}\le p\le\frac{N+q}{2}.
\end{equation} 
We further define the \emph{restricted fortuity window} by excluding its two boundaries,
\begin{equation}
    \label{eq:restrictedwindow}
    \frac{N-q}{2} < p < \frac{N+q}{2}.
\end{equation} 
When the boundary degrees are integral, additional BPS states may occur at the edges of the full fortuity window. Writing
$a=(N-q)/2$, exterior multiplication by the coupling form
$C\in\Lambda^q(\CC^N)$ defines
\begin{equation}
  M\colon\HH^a\longrightarrow\HH^{a+q}.
\end{equation}
The two spaces have equal dimension, and the top-form pairing identifies
$\HH^{a+q}\cong(\HH^a)^*$. For odd $q$, the resulting square matrix obeys
\begin{equation}
  M^T=(-1)^aM.
\end{equation}

For example, at $q=3$ and $N=5$, one has $a=1$ and
\begin{equation}
  \Lambda^1(\CC^5)\xrightarrow{\;C\wedge\;}\Lambda^4(\CC^5).
\end{equation}
The map is represented by a $5\times5$ antisymmetric matrix, which generically
has rank four. Its one-dimensional kernel gives one BPS state at each
boundary, $p=1$ and $p=4$.

The two boundary multiplicities are equal, but their fermion parities are opposite, so they cancel in the Witten index. Following~\cite{KanazawaWettig}, we call these paired boundary states \emph{exceptional}; they are shown as white squares in Figure~\ref{fig:bpsdata}.

\section{Fortuitous and monotonous BPS states}
\label{sec:fortuity}

Given the BPS spectrum for $N$ fermions, a natural question is how it changes when one additional fermion is introduced. Following~\cite{CCSY}, we relate the cohomologies at $N$ and $N+1$ through a long exact sequence. Its connecting homomorphism determines whether a class admits a charge-preserving lift to the enlarged theory. For clarity, we present the construction for $q=3$.

\subsection{Adding one fermion}
\label{sec:hilbert-grow}

Passing from $N$ to $N+1$ adds one fermionic mode
$\psi_{N+1}$. A state of fermion number~$p$ in the enlarged
Hilbert space has the unique decomposition
\begin{equation}\label{eq:state-decomp}
  \ket{\Psi}
  =
  \alpha+\psi_{N+1}\beta,
  \qquad
  \alpha\in\HH_N^p,
  \qquad
  \beta\in\HH_N^{p-1},
\end{equation}
where the first and second terms have the new mode unoccupied
and occupied, respectively. Thus
\begin{equation}\label{eq:Hp-split}
  \HH_{N+1}^p=\underbrace{\HH_N^p}_{\text{$\psi_{N+1}$
  unoccupied}}\;\oplus\;
  \underbrace{\psi_{N+1}\cdot\HH_N^{p-1}}_{\text{$\psi_{N+1}$
  occupied}}.
\end{equation}
We suppress the vacuum of the added mode throughout.

For $q=3$, the enlarged supercharge can be written as
\begin{equation}\label{eq:QN+1-split}
  Q_{N+1}
  =
  Q_N+C'\psi_{N+1},
  \qquad
  C'
  :=
  \sum_{1\leq i<j\leq N}
  C_{ij,N+1}\psi_i\psi_j.
\end{equation}
Here $Q_N$ contains only the old modes, while the two-fermion
operator $C'$ contains the couplings involving the new one.
Since $Q_N$ has odd fermion number, it anticommutes with
$\psi_{N+1}$, whereas $C'$ commutes with it. Moreover,
$\psi_{N+1}^2=0$. It follows that
\begin{equation}\label{eq:QN+1-action}
  Q_{N+1}\bigl(\alpha+\psi_{N+1}\beta\bigr)
  =
  Q_N\alpha
  +
  \psi_{N+1}\bigl(C'\alpha-Q_N\beta\bigr).
\end{equation}
Equivalently, relative to the decomposition
\eqref{eq:Hp-split},
\begin{equation}\label{eq:block-matrix}
  Q_{N+1}
  =
  \begin{pmatrix}
    Q_N & 0 \\[3pt]
    C'  & -Q_N
  \end{pmatrix}.
\end{equation}

The state $\alpha+\psi_{N+1}\beta$ is therefore
$Q_{N+1}$-closed precisely when
\begin{equation}\label{eq:closure-conditions}
  Q_N\alpha=0,
  \qquad
  C'\alpha=Q_N\beta.
\end{equation}
Thus $\alpha$ must be a cocycle of the smaller theory, and its
extension to the enlarged theory is obstructed only by the
cohomology class of $C'\alpha$. These statements are organized
by the long exact sequence described below.

\subsection{The long exact sequence}
\label{sec:LES}

The decomposition~\eqref{eq:Hp-split} and the block form~\eqref{eq:block-matrix} give a short exact sequence
\begin{equation}\label{eq:SES}
  0 \longrightarrow \psi_{N+1}\,\HH_N^{p-1}
  \xrightarrow{\;\;\iota\;\;}
  \HH_{N+1}^p
  \xrightarrow{\;\;\pi\;\;}
  \HH_N^p \longrightarrow 0,
\end{equation}
where $\iota$ is the inclusion of the occupied sector and $\pi$ projects onto the unoccupied sector. 

The short exact sequence~\eqref{eq:SES} induces the long
exact sequence in cohomology~\cite{CCSY}
\begin{equation}\label{eq:LES-concrete}
  \cdots
  \xrightarrow{\;\;\pi_*\;\;}
  H^{p-3}(Q_N)
  \xrightarrow{\;\delta\;}
  H^{p-1}(Q_N)
  \xrightarrow{\;\;\iota_*\;\;}
  H^p(Q_{N+1})
  \xrightarrow{\;\;\pi_*\;\;}
  H^p(Q_N)
  \xrightarrow{\;\;\delta\;\;}
  H^{p+2}(Q_N)
  \xrightarrow{\;\;\iota_*\;\;}\cdots
\end{equation} 
The induced maps are
\begin{equation}
  \pi_*[\alpha+\psi_{N+1}\beta]=[\alpha],
  \qquad
  \iota_*[\beta]=[\psi_{N+1}\beta],
  \qquad
  \delta[\alpha]=[C'\alpha].
\end{equation}
The connecting homomorphism $\delta$ measures the obstruction to a
charge-preserving lift: by exactness,
$  \im\pi_*=\ker\delta$.
Thus a class $[\alpha]\in H^p(Q_N)$ admits such a lift precisely when
$\delta[\alpha]=0$.

\begin{figure}[tb]
\begin{center}
\begin{tikzpicture}
\node[anchor=south west,inner sep=0] (img) at (0,0)
{\includegraphics[width=12.5cm]{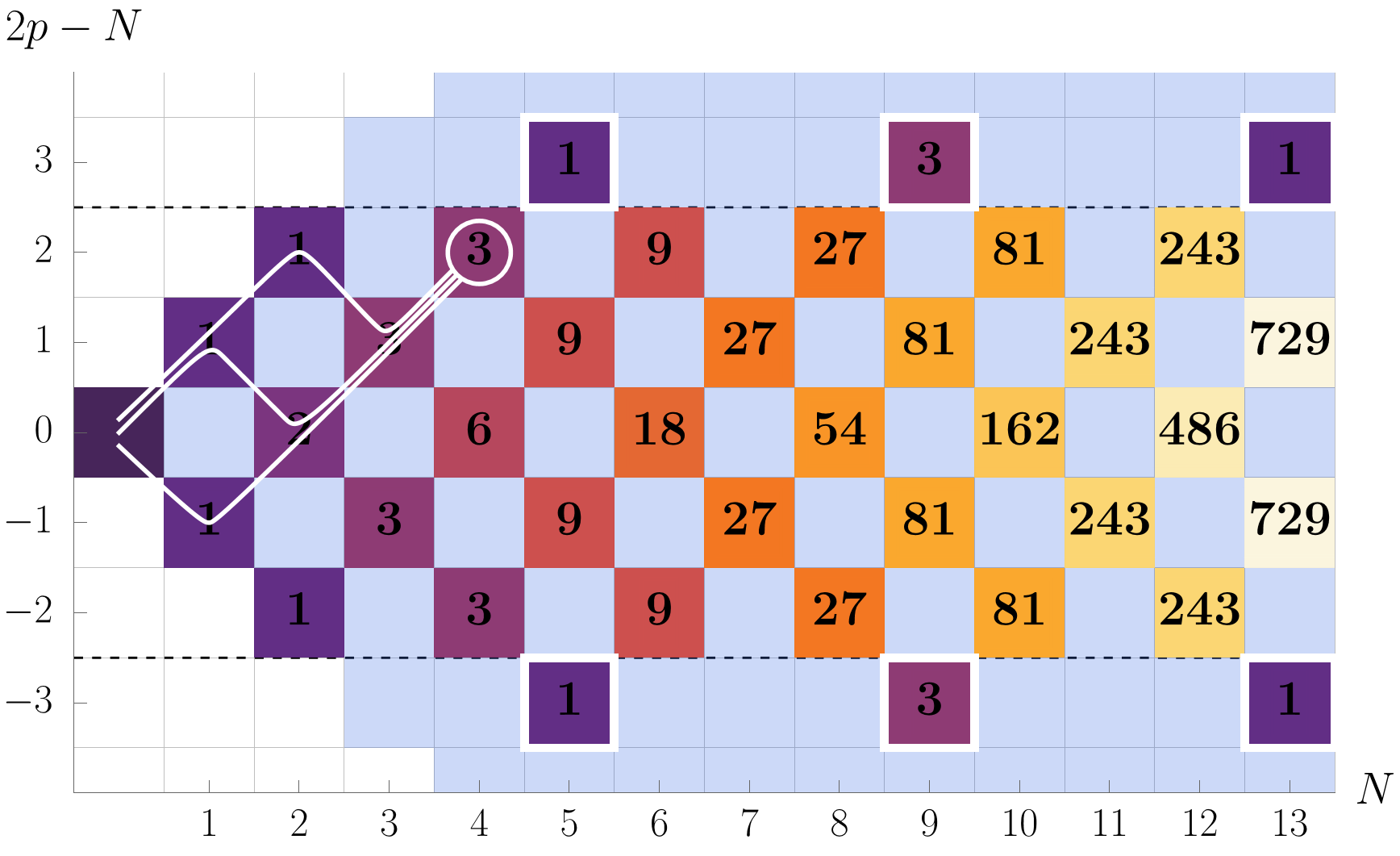}};

\node[rotate=90] at (13,3.75) {\scalebox{1.3}{$\underbrace{\hskip 3cm}_{\text{restricted fortuity window}}$}};

\end{tikzpicture}
\caption{\label{fig:bpsdatare}
The BPS data of Figure~\ref{fig:bpsdata}, shown in the $(2p-N,N)$ plane. The three highlighted walks correspond to the three BPS states at $(N,p)=(4,3)$.}
\end{center}
\end{figure}
 
\begin{figure}[tb]
\begin{center}
\begin{tikzpicture}
\node[anchor=south west,inner sep=0] (img) at (0,0)
{\includegraphics[width=11cm]{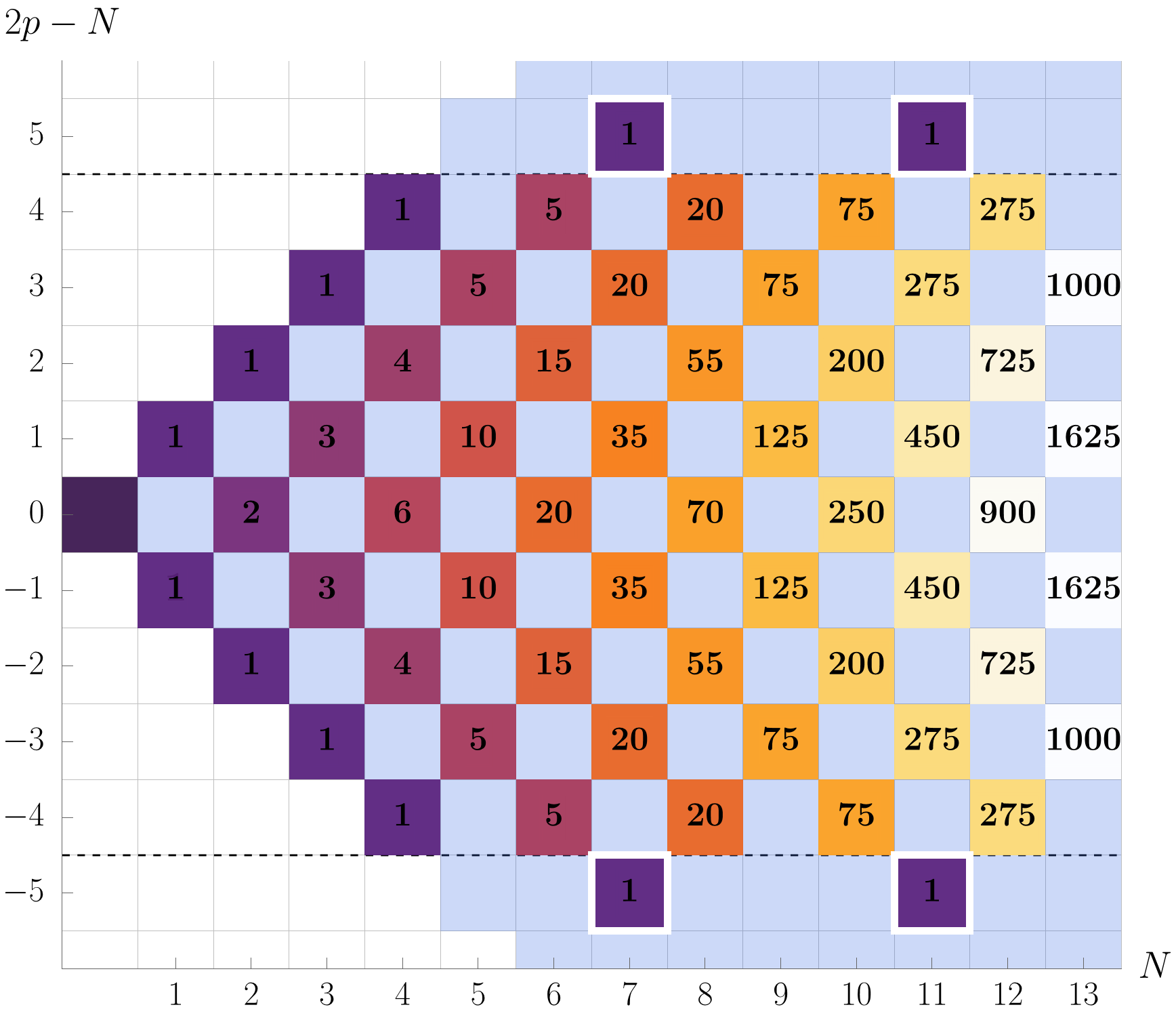}};

\node[rotate=90] at (11.5,4.75) {\scalebox{1.15}{$\underbrace{\hskip 5.5cm}_{\text{restricted fortuity window}}$}}; 

\end{tikzpicture}
\caption{\label{fig:bpsdatare5}
Dimensions of $H^p(Q_N)$ in the $\mathcal N=2$ supersymmetric SYK model with $q=5$ and generic couplings, displayed in the shifted $(2p-N,N)$ plane.}
\end{center}
\end{figure}

Exactness of the sequence yields the dimension formula
\begin{equation}\label{eq:dim-formula}
  \dim H^p(Q_{N+1})
  =\underbrace{\dim\ker(\delta|_{H^p})}_{\text{classes at $N$
  that lift}}
  +\underbrace{\dim\coker(\delta|_{H^{p-3}})}_{\text{classes from the occupied sector}},
\end{equation}
where $\coker(\delta|_{H^{p-3}})=H^{p-1}(Q_N)/\im(\delta|_{H^{p-3}})$.
When the two connecting homomorphisms entering
\eqref{eq:dim-formula} vanish, the dimension formula reduces to
\begin{equation}
  \dim H^p(Q_{N+1})
  =
  \dim H^p(Q_N)
  +
  \dim H^{p-1}(Q_N).
  \label{eq:dim-trivial-delta}
\end{equation}
Thus the BPS classes at degree $p$ in the enlarged theory arise from the
unoccupied sector at degree $p$ and the occupied sector built from degree
$p-1$ in the smaller theory. This recursion accounts for the BPS data in
Figures~\ref{fig:bpsdatare} and \ref{fig:bpsdatare5}, up to the exceptional boundary states.

\subsection{Classification of BPS states}
\label{sec:fort-mono-def}

A BPS class is \emph{monotonous} (or monotone) if it descends from a supersymmetric class
of the large-$N$ theory, and \emph{fortuitous} if it exists only over a finite range of system sizes~\cite{ChangLin}.

In the generic $\mathcal N=2$ SYK model, the BPS cohomology is confined to
the window~\eqref{eq:window}.
At fixed fermion number~$p$, this window eventually moves beyond the chosen charge sector as $N$ increases, after which $H^p(Q_N)$ vanishes. Consequently, every BPS class in the generic SYK model is fortuitous~\cite{CCSY}.

\section{Cohomological walks}
\label{sec:walks}

The long exact sequence~\eqref{eq:LES-concrete} relates the BPS cohomologies
at system sizes $N$ and $N+1$. In this section, we iterate its two uplift
channels and encode their successive choices by lattice walks, or binary
strings. The resulting walk labels are not canonical: they depend on choices
of representatives and splittings of the exact sequence. Nevertheless, we
show that every nonzero BPS class admits an uplift to arbitrarily
large~$N$ along at least one walk.

A walk of length~$N$ is a binary string
\begin{equation}
  w=w_1w_2\cdots w_N\in\{0,1\}^N.
\end{equation}
The bit $w_k$ records the channel chosen when the $k$-th fermion is added:
$w_k=0$ denotes the unoccupied channel (``down step''), while $w_k=1$ denotes the
occupied channel (``up step''). These labels specify the distinguished component
used in the uplift; a dressed closed representative may have support
in both occupation sectors. The
sum $\sum_{k=1}^N w_k$ equals the fermion
number~$p$. For a sector $(N,p)$, define the \emph{level} by 
\begin{equation} 
    \ell:=2p-N. 
\end{equation} 
In these variables, the fortuity and restricted fortuity windows become $|\ell|\leq q$ and $|\ell|\leq q-1$, respectively.

\subsection{The two uplift channels}
\label{ss:walk-construction}

The long exact sequence may be iterated from one system size to the next.
Suppose first that $[\alpha]\in H^p(Q_{n-1})$ lies in the kernel of the
outgoing connecting homomorphism. It then admits an unoccupied sector uplift
to $H^p(Q_n)$,
\begin{equation}
    \label{eq:walk-down} [\alpha] \longmapsto [\alpha+\psi_n\beta], 
    \qquad 
    Q_{n-1}\beta=C'_{n-1}\alpha. 
\end{equation} 
This uplift need not be unique. Shifting $\beta$ by a
$Q_{n-1}$-closed cochain generally changes the cohomology class upstairs
while leaving its projection equal to $[\alpha]$. Thus the parent class may
have several possible uplifts through the unoccupied sector.

The occupied sector uplift is induced by
\begin{equation}\label{eq:walk-up}
  [\alpha]
  \longmapsto 
  [\psi_n\alpha].
\end{equation}
By exactness, this class is nonzero precisely when $
  [\alpha]\notin\operatorname{im}\delta$,
where $\delta$ is the connecting homomorphism into
$H^p(Q_{n-1})$.  

A \emph{cohomological walk} is a sequence of nonzero BPS classes related at
each step by one of these two uplift channels. A down step preserves the
fermion number and changes $\ell=2p-n$ by $-1$, while an up step raises the
fermion number by one and changes $\ell$ by $+1$. We encode these choices by a binary string $w\in\{0,1\}^n$, with $0$ and $1$
denoting the two uplift channels. Since the
uplifts may not be unique, a walk records the chosen
channels but does not by itself determine a unique sequence of classes. We
return to this ambiguity in Section~\ref{sec:obstructions}.

Appendix~\ref{app:walks} gives the counting interpretation:
the ordinary, non-exceptional BPS multiplicities equal the numbers of walks
confined to the restricted strip $|\ell|\leq q-1$. The exceptional boundary
classes lie at $\ell=\pm q$ and have to be treated separately.

\subsection{Uplifting to arbitrarily large $N$}
\label{ss:infinite}

We now ask whether a nonzero BPS class can always be uplifted to the next
system size when either channel is allowed. For a class in $H^p(Q_N)$, the
two relevant connecting homomorphisms are
\begin{align}
  \delta_{\mathrm{out}}
  &\colon H^p(Q_N)\longrightarrow H^{p+q-1}(Q_N),
  \label{eq:delta-out}\\
  \delta_{\mathrm{in}}
  &\colon H^{p-q+1}(Q_N)\longrightarrow H^p(Q_N).
  \label{eq:delta-in}
\end{align}
The first controls the unoccupied sector lift, while the image of the second
is the kernel of the occupied sector map.

The map $\delta_{\mathrm{out}}$ can be nonzero only if its target lies inside
the fortuity window. This requires
\begin{equation}
  \ell+2(q-1)\leq q,
  \qquad\text{or equivalently}\qquad
  \ell\leq2-q,
\end{equation}
where $\ell=2p-N$.
Similarly, $\delta_{\mathrm{in}}$ can be nonzero only if its source lies
inside the window, which requires
\begin{equation}
  \ell-2(q-1)\geq-q,
  \qquad\text{or equivalently}\qquad
  \ell\geq q-2.
\end{equation}
These inequalities are necessary conditions for the corresponding connecting homomorphisms to be nonzero. For odd $q\geq3$, the two conditions are incompatible, so the two connecting homomorphisms can never be nonzero simultaneously.
Thus every nonzero BPS class has
at least one nonzero uplift  to the next system size.  

Iterating the argument shows that every nonzero BPS class can be uplifted
to arbitrarily large~$N$ along a lattice walk contained in the full fortuity
window. At the boundary $\ell=+q$, only the down step remains available,
while at $\ell=-q$ only the up step remains available.

This gives a new perspective on fortuity. At fixed fermion number, every BPS
class in generic supersymmetric SYK eventually leaves the moving window and
is fortuitous. Nevertheless, if its charge is allowed to change, the class
always has a way forward. The same finite width that causes fortuity at fixed
charge prevents the two uplift channels from being obstructed
simultaneously.

\subsection{Obstructions to a canonical uplift}
\label{sec:obstructions}

The walks uplift BPS cohomology classes. To track
physical BPS states, however, one must select their harmonic representatives,
annihilated by both $Q$ and $Q^\dagger$. We now describe two independent
sources of ambiguity: an algebraic uplift need not preserve harmonicity, and
the uplifted cohomology class itself need not be unique. Resolving either
ambiguity requires the metric data of the enlarged theory.

First consider harmonicity. Let
$\alpha\in\ker H_N\cap\HH_N^p$. The block form
\eqref{eq:block-matrix} is lower triangular, while its adjoint is upper
triangular. For the down channel, the bare embedding  $\binom{\alpha}{0}$ satisfies
\begin{equation}\label{eq:down-closures}
  Q_{N+1}
  \binom{\alpha}{0}
  =
  \binom{0}{C'\alpha},
  \qquad
  Q_{N+1}^\dagger
  \binom{\alpha}{0}
  =
  \binom{0}{0}.
\end{equation}
It is therefore automatically $Q_{N+1}^\dagger$-closed, but generically not
$Q_{N+1}$-closed. If $[C'\alpha]=0$, a correction $\beta$ satisfying
\begin{equation}
  Q_N\beta=C'\alpha
\end{equation}
restores $Q_{N+1}$-closure. The corrected representative nevertheless obeys
\begin{equation}
  Q_{N+1}^\dagger
  \binom{\alpha}{\beta}
  =
  \binom{(C')^\dagger\beta}{-Q_N^\dagger\beta},
\end{equation}
which need not vanish. Thus the cohomological correction does not in general
produce the harmonic representative.

For the up channel, the roles are reversed:
\begin{equation}\label{eq:up-closures}
  Q_{N+1}
  \binom{0}{\alpha}
  =
  \binom{0}{0},
  \qquad
  Q_{N+1}^\dagger
  \binom{0}{\alpha}
  =
  \binom{(C')^\dagger\alpha}{0}.
\end{equation}
The bare state is automatically $Q_{N+1}$-closed, but generically not
$Q_{N+1}^\dagger$-closed. In either channel, obtaining a physical BPS state
requires projection onto the harmonic subspace of the enlarged theory and
may therefore modify both occupation components.

A second ambiguity already arises at the level of cohomology. If
$[\alpha]\in\ker\delta\subset H^p(Q_N)$, exactness gives at least one class
in $H^p(Q_{N+1})$ that projects to $[\alpha]$. Any two such classes differ
by an element of $\ker\pi_*=\im\iota_*$. Hence the fiber is the affine coset
\begin{equation}\label{eq:coset}
  \pi_*^{-1}\bigl([\alpha]\bigr)
  =
  [\widetilde\alpha]+\im\iota_*,
\end{equation}
where $[\widetilde\alpha]$ is any chosen uplift. Distinct elements of this
fiber are distinct cohomology classes upstairs and may have different
uplift properties at the next system size. The binary walk records the
sequence of uplift channels, but does not by itself select a unique class
along that walk.

These ambiguities explain why cohomological uplift does not canonically
determine an uplift of physical BPS states. Hodge representatives depend on
the adjoint and harmonic projector, which are not functorial under the
algebraic maps in the long exact sequence. A chain map need not commute with
adjoints, Hamiltonians, or harmonic projectors. Selecting an uplifted class
and its harmonic representative must therefore use the Hilbert-space metric
and the coupling-dependent supercharge of the enlarged theory.

\section{The uplift decoder}
\label{sec:rayleigh}

The cohomological construction does not determine whether a harmonic state at
size $N$ occurs as a definite component of a harmonic state in the enlarged
theory. It also does not measure how much additional dressing is required when
such an uplift exists. Both questions are encoded in a single linear map,
the uplift decoder.

\subsection{Decoder and exact uplift}
\label{sec:rayleigh-def}

\begin{figure}[t]
  \centering
  \includegraphics[width=0.50\linewidth]{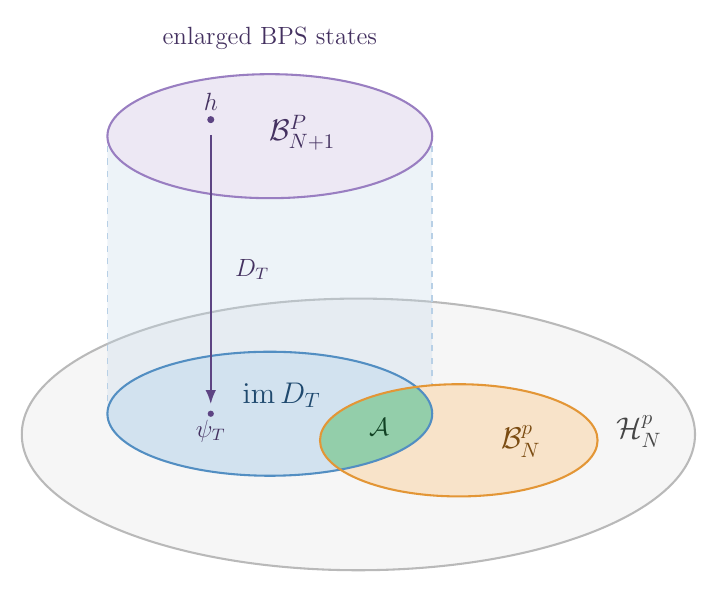}
  \caption{The decoder as a projection. An enlarged BPS state
    $h\in\mathcal B_{N+1}^{P}$ is projected onto its charge-$T$ slot,
     landing in the image $\operatorname{im}D_T\subseteq\mathcal H_N^{p}$.
    A source state lifts exactly when it lies in the intersection
    $\mathcal A=\operatorname{im}D_T\cap\mathcal B_N^{p}$ (green); outside it the
    lift is only approximate.}
  \label{fig:decoder-concept}
\end{figure}

Let $M$ be the set of fermionic modes added in passing from size $N$ to
$N'$. A state of total fermion number $P$ decomposes according to the
occupation pattern of the added modes,
\begin{equation}
  \ket{\Psi}
  =
  \sum_{S\subseteq M}\alpha_S\otimes\ket{S},
  \qquad
  \alpha_S\in\HH_N^{P-|S|},
  \label{eq:general-decomp}
\end{equation}
where $\ket{S}$ occupies precisely the modes in $S$. Fix a channel
$T\subseteq M$ and a charge $p$, so that $P=p+|T|$. The component map
$\Pi_T\ket{\Psi}=\alpha_T$, restricted to the target BPS space, defines the
\emph{decoder}
\begin{equation}
  D_T
  :=
  \Pi_T\big|_{\mathcal B_{N'}^P}
  :
  \mathcal B_{N'}^P
  \longrightarrow
  \HH_N^p. 
  \label{eq:decoder}
\end{equation}
Its image consists of the states downstairs that occur as the exact
$T$-component of a BPS state upstairs (Figure~\ref{fig:decoder-concept}).

For a nonzero $\alpha\in\mathcal B_N^p$, an exact uplift through the channel
$T$ exists if and only if
\begin{equation}
  \alpha\in\im D_T,
  \label{eq:exact-lift-criterion}
\end{equation}
or equivalently if there exists $h\in\mathcal B_{N'}^P$ such that $D_Th=\alpha$.  The \emph{exact-lift locus} is therefore
\begin{equation}\label{eq:exact-lift-locus}
  \mathcal A^p_{N;T}:=\mathcal B_N^p\cap\im D_T .
\end{equation}
For ordinary SYK with one added fermion, the two channels are
$T=\varnothing$, the down channel, and $T=\{\psi_{N+1}\}$, the up channel.

\subsection{The decoder spectral measure}
\label{sec:decoder-measure}

The positive operator
\begin{equation}
  K_T:=D_TD_T^\dagger
  \label{eq:decoder-kernel}
\end{equation}
acts on $\HH_N^p$ and contains metric information relevant to the
uplift. Write the singular-value decomposition of the decoder as
\begin{equation}
  D_T
  =
  \sum_{\sigma_j>0}
  \sigma_j\ket{u_j}\bra{v_j},
  \label{eq:decoder-svd}
\end{equation}
where the $\ket{v_j}\in\mathcal B_{N'}^P$ are orthonormal right singular
vectors, the $\ket{u_j}\in\HH_N^p$ are orthonormal left singular vectors, and
$\sigma_j>0$ are the nonzero singular values. It follows that
\begin{equation}
  K_T
  =
  \sum_{\sigma_j>0}
  \sigma_j^2\ket{u_j}\bra{u_j}.
\end{equation}
Since $D_T$ is the restriction of an orthogonal component projection, the eigenvalues of $K_T$ are
$  \lambda_j=\sigma_j^2\in(0,1]$.

Let $\{u_j\}$ be a complete orthonormal eigenbasis of $K_T$, including its
zero eigenspace, and expand the normalized state
\begin{equation}
  \widehat\alpha
  :=
  \frac{\alpha}{\|\alpha\|}
  =
  \sum_j a_j u_j.
\end{equation}
The decoder spectral measure seen by $\alpha$ is
\begin{equation}
  d\mu_{\alpha,T}(\lambda)
  :=
  \sum_j |a_j|^2\,
  \delta(\lambda-\lambda_j)\,d\lambda,
  \qquad
  \int_{[0,1]}d\mu_{\alpha,T}(\lambda)=1.
  \label{eq:decoder-measure}
\end{equation}
The mass at $\lambda=0$ is the part of $\alpha$ invisible to the decoder.
Directions with $0<\lambda\ll1$ are visible but require a large amplification
upstairs, while directions with $\lambda=1$ pass through the decoder without
dressing.

\subsection{Fidelity, bare fraction, and dressing cost}
\label{sec:dressing}
\label{sec:alignment}

We define the uplift fidelity as the total visible weight,  
\begin{equation}
  \cF_T(\alpha)
  :=
  \int_{\lambda>0}d\mu_{\alpha,T}(\lambda)
  =\sum_{\lambda_j>0}|a_j|^2 =
  \frac{\bigl\|P_{\im D_T}\alpha\bigr\|^2}{\|\alpha\|^2}
  \in[0,1].
  \label{eq:uplift-fidelity}
\end{equation}
Thus
\begin{equation}
  \cF_T(\alpha)=1
  \iff
  \alpha\in\im D_T,
\end{equation}
whereas $\cF_T(\alpha)<1$ means that only the visible projection $ \alpha_\parallel
  :=
  P_{\im D_T}\alpha$
can be reconstructed through this channel.

The \emph{bare embedding} through channel~$T$ is obtained without adding components in any other sector: 
\begin{equation} 
    E_T\alpha := \alpha\otimes\ket{T}. 
    \label{eq:bare-embedding} 
\end{equation}
Since $E_T=\Pi_T^\dagger$ and $D_T=\Pi_T|_{\mathcal B_{N'}^P}$, the adjoint decoder satisfies 
\begin{equation} 
    D_T^\dagger\alpha = P_{\mathcal B_{N'}^P}E_T\alpha.
    \label{eq:adjoint-decoder-bare} 
\end{equation}
We define the \emph{bare fraction} as the first moment of the decoder spectral measure,  
\begin{equation}
  f_{0,T}(\alpha)
  :=
  \int_{\lambda>0}\lambda\,d\mu_{\alpha,T}(\lambda).
  \label{eq:bare-fraction-moment}
\end{equation}
Using the singular-value decomposition and
$\widehat\alpha=\sum_j a_j\ket{u_j}$, we have
\begin{equation}
  \frac{D_T^\dagger\alpha}{\|\alpha\|}
  =
  \sum_{\lambda_j>0}
  \sqrt{\lambda_j}\,a_j\ket{v_j},
\end{equation}
and hence, by orthonormality of the right singular vectors,
\begin{equation}
  f_{0,T}(\alpha)
  =
  \sum_{\lambda_j>0}\lambda_j|a_j|^2
  =
  \frac{\|D_T^\dagger\alpha\|^2}{\|\alpha\|^2}
  =
  \frac{
    \bigl\|P_{\mathcal B_{N'}^P}E_T\alpha\bigr\|^2
  }{\|\alpha\|^2}.
  \label{eq:bare-fraction-projection}
\end{equation}
Thus $f_{0,T}$ is the fraction of the bare embedding's squared norm contained in the enlarged BPS subspace.
Since $0\leq\lambda\leq1$,
\begin{equation}
  0\leq f_{0,T}(\alpha)\leq\cF_T(\alpha)\leq1.
  \label{eq:f0-fidelity-bound}
\end{equation}

The inverse moment measures the norm amplification required to reconstruct the
visible component,
\begin{equation}
  m_{-1,T}(\alpha)
  :=
  \int_{\lambda>0}\frac{1}{\lambda}\,
  d\mu_{\alpha,T}(\lambda).
  \label{eq:inverse-moment}
\end{equation}
Using the singular-value decomposition, the Moore--Penrose
pseudoinverse of $D_T$ acts as
\begin{equation}
  \frac{D_T^+\alpha}{\|\alpha\|}
  =
  \sum_{\lambda_j>0}
  \frac{a_j}{\sqrt{\lambda_j}}\ket{v_j}.
\end{equation}
By orthonormality of the right singular vectors, it follows that
\begin{equation}
  m_{-1,T}(\alpha)
  =
  \sum_{\lambda_j>0}\frac{|a_j|^2}{\lambda_j}
  =
  \frac{\|D_T^+\alpha\|^2}{\|\alpha\|^2}.
  \label{eq:inverse-moment2}
\end{equation}

When $\cF_T(\alpha)=1$, the state has an exact uplift and $D_T^+\alpha$
is its unique minimum-norm uplift. Its dressing cost is
\begin{equation}
  \kappa_T(\alpha)
  :=
  \frac{\|D_T^+\alpha\|^2}{\|\alpha\|^2}-1
  =
  m_{-1,T}(\alpha)-1
  \geq0.
  \label{eq:kappa-def}
\end{equation}
It vanishes precisely when the bare embedding
$E_T\alpha=\alpha\otimes\ket{T}$ is already BPS.

When $0<\cF_T(\alpha)<1$, the pseudoinverse reconstructs
$\alpha_\parallel$ rather than the full state. The corresponding projected
dressing cost is
\begin{equation}
  \kappa_{\mathrm{proj},T}(\alpha)
  :=
  \frac{\|D_T^+\alpha\|^2}{\|\alpha_\parallel\|^2}-1
  =
  \frac{m_{-1,T}(\alpha)}{\cF_T(\alpha)}-1
  \geq0.
  \label{eq:kappa-proj}
\end{equation}
This is the cost of uplifting the normalized visible projection, not the
original state. On the exact-lift locus it reduces to the ordinary dressing
cost.

The three diagnostics are therefore moments of the same spectral measure:
\begin{equation}
  \cF_T(\alpha)
  =
  \int_{\lambda>0}d\mu_{\alpha,T},
  \qquad
  f_{0,T}(\alpha)
  =
  \int_{\lambda>0}\lambda\,d\mu_{\alpha,T},
  \qquad
  m_{-1,T}(\alpha)
  =
  \int_{\lambda>0}\lambda^{-1}d\mu_{\alpha,T}.
  \label{eq:decoder-moments-summary}
\end{equation}
They measure, respectively, the visible fraction of the state, the fraction of the bare embedding already contained in the enlarged BPS space, and how much norm increase is needed to reconstruct the visible part.

\section{Solvable models}
\label{sec:solvable}

The last section formulated the metric uplift problem in terms of the decoder and its spectral diagnostics. We now evaluate these quantities in two analytically controlled settings that realize perfect uplift through different mechanisms. In the single-matrix model of Chen~\cite{Chen}, we identify a bare embedding under which every BPS state remains harmonic, giving exact uplift with no dressing.
In the symmetrized two-flavor model of~\cite{CCSY}, a conserved bilinear gives every tower state an exact, explicitly dressed uplift.

\subsection{The single-matrix model}
\label{sec:chen-model}

An especially rigid uplift mechanism occurs in the single-matrix model of
Chen~\cite{Chen}. The degrees of freedom are complex fermions arranged into an
$N\times N$ matrix $\Psi$, with supercharge
\begin{equation}
  Q_N=\Tr\Psi^3,
  \qquad
  H_N=\{Q_N,Q_N^\dagger\}.
  \label{eq:chen-Q}
\end{equation}
The trace mode decouples, so we work throughout with the $N^2-1$ traceless
modes. 

Let $T^a$, $a=1,\ldots,N^2-1$, be Hermitian generators of $\mathfrak{su}(N)$ in the fundamental representation, normalized by 
\begin{equation} 
\Tr(T^aT^b)=\frac12\delta^{ab}, 
\qquad     
[T^a,T^b]=if^{abc}T^c. 
\end{equation} 
Expanding 
\begin{equation} 
    \Psi = \frac{\psi_0}{\sqrt N}\,\mathbf 1 +\sqrt{2}\,\psi^aT^a, 
\end{equation} 
the terms containing $\psi_0$ vanish by fermionic antisymmetry. Only the fully antisymmetric part of $\Tr(T^aT^bT^c)$
contributes, giving
\begin{equation}
  Q_N
  =
  \frac{i}{\sqrt{2}}\,
  f^{abc}\psi^a\psi^b\psi^c.
  \label{eq:chen-Qadj}
\end{equation}
The Hamiltonian is determined by the quadratic Casimir of the global $SU(N)$
action,
\begin{equation}
  J^a=-if^{abc}\psi^b(\psi^c)^\dagger,
  \qquad
  C_2=\sum_{a=1}^{N^2-1}J^aJ^a,
  \qquad
  H_N=3N(N^2-1)\mathbf 1-9C_2.
  \label{eq:chen-H}
\end{equation}
The BPS states therefore form copies of the maximal-Casimir staircase
representation appearing in the fermionic Fock space~\cite{Chen}.

A highest weight state is obtained by filling the strictly lower triangle,
\begin{equation}
  \ket{\lambda_N}
  =
  \prod_{1\leq j<i\leq N}\Psi_{ij}\ket{0},
  \label{eq:chen-highest-weight}
\end{equation}
with the factors taken in a fixed order. The BPS space consists of copies of
the maximal-Casimir staircase representation generated from
$\ket{\lambda_N}$ and its decorations by subsets of the $N-1$ zero-weight
Cartan fermions. In the traceless sector, its fermion-number support is
\begin{equation}
  \frac{N(N-1)}{2}
  \leq p\leq
  \frac{N(N+1)}{2}-1.
  \label{eq:chen-window}
\end{equation}
The decoupled trace fermion supplies one further optional zero-weight
decoration. Restoring it extends the upper endpoint by one.
Because these windows move upward with $N$, every BPS state at fixed fermion
number is fortuitous.

Under the embedded subgroup $SU(N)\subset SU(N+1)$, the enlarged adjoint
branches as
\begin{equation}
  \operatorname{adj}_{SU(N+1)}
  =
  \operatorname{adj}_{SU(N)}
  \oplus\mathbf N
  \oplus\overline{\mathbf N}
  \oplus\mathbf 1.
  \label{eq:chen-adjoint-branching}
\end{equation}
Thus increasing the rank adds a new column, a new row, and one additional
traceless Cartan mode. Let $\ket{0}_{\mathrm{ext}}$ denote the Fock vacuum of
these new modes. With the lower-triangular convention of
\eqref{eq:chen-highest-weight}, the natural staircase channel fills the new row:
\begin{equation}
  T_{\mathrm{row}}
  :=
  \prod_{j=1}^{N}\Psi_{N+1,j}\ket{0}_{\mathrm{ext}}.
  \label{eq:chen-row-fill}
\end{equation}
The new row modes transform in the antifundamental representation of $SU(N)$,
so their completely antisymmetric product transforms by $\det(U^\dagger)=1$.
Thus $T_{\mathrm{row}}$ is an $SU(N)$ singlet. With the standard
Fock-space normalization, $\|T_{\mathrm{row}}\|=1$. Consequently, the bare staircase embedding
\begin{equation}
  E_N(\alpha):=\alpha\otimes T_{\mathrm{row}}
  \label{eq:chen-embedding}
\end{equation}
is isometric and equivariant under the embedded subgroup,
\begin{equation}
  E_N(g\alpha)=gE_N(\alpha),
  \qquad g\in SU(N)\subset SU(N+1).
  \label{eq:chen-equivariance}
\end{equation}

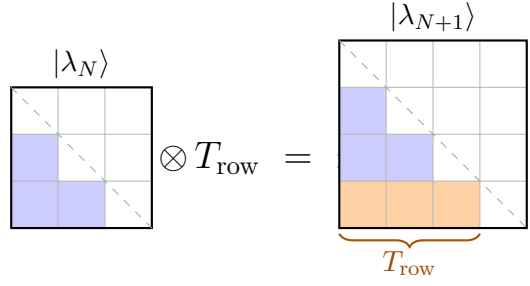
\begin{figure}[t]
\centering
\begin{tikzpicture}[scale=0.62, every node/.style={font=\small}]
\begin{scope}
  \fill[blue!20] (0,1) rectangle (1,2);
  \fill[blue!20] (0,0) rectangle (1,1);
  \fill[blue!20] (1,0) rectangle (2,1);
  \draw[gray!55, thin] (0,0) grid (3,3);
  \draw[thick] (0,0) rectangle (3,3);
  \draw[dashed, gray!70] (0,3) -- (3,0);
  \node at (1.5,3.5) {$\ket{\lambda_N}$};
\end{scope}

\node at (5.35,1.5) {\large $\otimes\,T_{\mathrm{row}}\ =\ \pm$};

\begin{scope}[shift={(7,0)}]
  \fill[blue!20] (0,2) rectangle (1,3);
  \fill[blue!20] (0,1) rectangle (1,2);
  \fill[blue!20] (1,1) rectangle (2,2);
  \fill[orange!35] (0,0) rectangle (1,1);
  \fill[orange!35] (1,0) rectangle (2,1);
  \fill[orange!35] (2,0) rectangle (3,1);
  \draw[gray!55, thin] (0,0) grid (4,4);
  \draw[thick] (0,0) rectangle (4,4);
  \draw[dashed, gray!70] (0,4) -- (4,0);
  \node at (2,4.5) {$\ket{\lambda_{N+1}}$};
  \draw[decorate, decoration={brace, amplitude=5pt, mirror},
        thick, orange!65!black] (0,-0.12) -- (3,-0.12);
  \node[orange!60!black] at (1.5,-0.75) {$T_{\mathrm{row}}$};
\end{scope}
\end{tikzpicture}
\caption{The staircase uplift for $N=3$. The filled new row $T_{\mathrm{row}}$ (orange)
completes the lower triangle, so
$\ket{\lambda_N}\otimes T_{\mathrm{row}}=\pm\ket{\lambda_{N+1}}$, therefore the state at $N+1$ is automatically BPS.}
\label{fig:phistar}
\end{figure}

Filling the new row completes the staircase,
\begin{equation}
  E_N\ket{\lambda_N}
  =
  \ket{\lambda_N}\otimes T_{\mathrm{row}}
  =
  \pm\ket{\lambda_{N+1}},
  \label{eq:chen-staircase-identity}
\end{equation}
where the sign depends only on the chosen ordering of the fermionic modes.
More generally, if $C_S$ is any product of embedded zero-weight traceless
Cartan fermions, then
\begin{equation}
  E_N\bigl(C_S\ket{\lambda_N}\bigr)
  =
  \pm C_S\ket{\lambda_{N+1}}.
  \label{eq:chen-decoration-uplift}
\end{equation}
The state on the right is a decorated
highest weight state of a maximal Casimir BPS multiplet at rank $N+1$.

Each corresponding rank-$N$ BPS multiplet is generated from
$C_S\ket{\lambda_N}$ by the lowering operators of $SU(N)$. If $X$ is a
product of such operators, equivariance gives
\begin{equation}
  E_N\bigl(XC_S\ket{\lambda_N}\bigr)
  =
  \pm XC_S\ket{\lambda_{N+1}}.
  \label{eq:chen-descendant-uplift}
\end{equation}
Since $C_S\ket{\lambda_{N+1}}$ is BPS and $H_{N+1}$ commutes with the
embedded $SU(N)$ action, the state on the right is also BPS. The decorated
multiplets span $\ker H_N$, and therefore
\begin{equation}
  E_N(\ker H_N)\subseteq\ker H_{N+1}.
  \label{eq:chen-bps-inclusion}
\end{equation}
Thus the bare staircase uplift of every rank-$N$ BPS state is already
harmonic at rank $N+1$.

For the filled-row channel $T_{\mathrm{row}}$, $E_N(\alpha)$ has
$T_{\mathrm{row}}$-component $\alpha$ and no components in the other new-mode sectors.
Since the occupation sectors are orthogonal and $E_N$ is isometric, this bare
BPS state is the unique minimum-norm exact uplift. We have,
\begin{equation}
  \Pi_{T_{\mathrm{row}}}E_N=\mathbf 1_{\ker H_N},
  \qquad
  E_N^\dagger E_N=\mathbf 1_{\ker H_N}.
  \label{eq:chen-decoder-identity}
\end{equation}
The decoder diagnostics consequently take the rigid values
\begin{equation}
  \cF_{T_{\mathrm{row}}}(\alpha)=1,
  \qquad
  f_{0,T_{\mathrm{row}}}(\alpha)=1,
  \qquad 
  \kappa_{T_{\mathrm{row}}}(\alpha)=0
  \label{eq:chen-profile}
\end{equation}
for every BPS state $\alpha$.

There is a second staircase channel. Let $\eta_{N+1}$ denote the normalized additional
zero-weight traceless Cartan fermion in
\eqref{eq:chen-adjoint-branching}, which is an $SU(N)$ singlet, and define
\begin{equation}
  T_{\mathrm{row+C}}
  :=
  \eta_{N+1}T_{\mathrm{row}}.
  \label{eq:chen-cartan-row-fill}
\end{equation}
The rank-$(N+1)$ BPS classification implies that
$\eta_{N+1}C_S\ket{\lambda_{N+1}}$ is another decorated maximal-Casimir
highest weight. Repeating the equivariance argument gives a second bare,
harmonic, and isometric embedding of the full BPS space.

The filled-row and Cartan-dressed channels shift the fermion number by
$N$ and $N+1$, respectively. 
Together they reach every fermion number in the rank-$(N+1)$ BPS window,
although this does not assert that their images exhaust the full BPS space
upstairs. Matrix transposition $\Psi\mapsto\Psi^{\mathsf T}$ exchanges the
 row and column constructions while preserving the Hamiltonian,
 providing two further correction-free column channels.

The single-matrix model is thus an exactly solvable skeleton of fortuity. Its
BPS states are fortuitous at fixed fermion number, but the metric obstruction
to uplift is absent: every BPS state admits the same bare isometric
uplift with unit fidelity and zero dressing cost.

\subsection{The two-flavor SYK model}
\label{sec:twoflavor}

The single-matrix uplift was lossless. The second solvable mechanism maintains fidelity at one through a genuinely dressed uplift, $\kappa_T>0$, and produces a monotonous family rather than a fortuitous one. The model is a supersymmetric SYK theory with two flavors of complex fermions
$\psi_i$ and $\chi_i$, with $i=1,\dots,N$. Its cubic supercharge has its last two indices symmetrized,
\begin{equation}
Q=\sum_{ijk}C_{ijk}\,\psi_i\psi_j\chi^\dagger_k ,\qquad H=\{Q,Q^\dagger\},
\label{eq:Qtf}
\end{equation}
with random couplings $C_{ijk}$ symmetric in $(j,k)$~\cite{CCSY, Heydeman:2022lse}.
The supercharge carries charge $(+2,-1)$ under the two flavor numbers $(\Npsi,\Nchi)$, whereas the Hamiltonian preserves them. Passing $N\to N+1$ adjoins one pair of modes
$(\psi_{N+1},\chi_{N+1})$, so the new occupation is $T=(t_\psi,t_\chi)\in\{0,1\}^2$: the four
channels $00,\ 10,\ 01,\ 11$ (charge-preserving, $+\psi$, $+\chi$, $+\psi{+}\chi$) are the concrete
directions of the uplift. Thus, for a source BPS state $\alpha\in\Bps_N^{(a,b)}$, channel $T$ targets the BPS sector $\Bps_{N+1}^{(A,B)}$ with charge $(A,B)=(a+t_\psi,b+t_\chi)$.
 
The supercharge commutes with the fermion bilinear
\begin{equation}
V=\sum_{i}\psi_i\chi_i,\qquad [Q,V]=0 .
\label{eq:Vtf}
\end{equation}
This conserved bilinear generates a monotonous tower of BPS states~\cite{CCSY},
\begin{equation}
\ket{V^n}=V^n\ket\Omega,\qquad n=0,1,\dots,N,
\label{eq:tower}
\end{equation}
at charge $(n,n)$ and fermion number $2n$. The tower contains only $\mathcal{O}(N)$ states, in contrast to the much larger collection of fortuitous BPS states. Throughout, ``fortuitous''
means the orthogonal complement of the tower inside $\Bps_N$ at a given charge.

The uplift from $N$ to $N+1$ enlarges the bilinear by a single term, $V_{N+1}=V_N+\psi_{N+1}\chi_{N+1}$. This term squares to
zero and commutes with $V_N$. Hence, the binomial expansion of $V_{N+1}^n$ stops after one step:
\begin{equation}
V_{N+1}^{\,n}\ket{\Omega_{N+1}}=\ket{V_N^n}\otimes\ket{00}
\;+\;n\,\ket{V_N^{n-1}}\otimes\ket{11}.
\label{eq:lift}
\end{equation}
The first term is the bare embedding of the rung-$n$ tower state into the empty-pair channel, while the
second is a correction supported entirely in the doubly-occupied ($11$) sector, the singly-occupied
corrections vanishing. The right-hand side is harmonic at $N+1$ and its $00$-component is exactly the
tower state below, so the decoder returns it and $\mathcal F_{00}(\ket{V^n})=1$ for all  $n\leq N$. By the same reasoning, $\mathcal F_{11}(\ket{V^n})=1$.

The explicit completion in \eqref{eq:lift} is genuinely dressed whenever $n>0$. The states are normalized as
\begin{equation} 
    \left\|\ket{V_N^n}\right\|^2 = \frac{n!\,N!}{(N-n)!}.
\end{equation} 
Since $\kappa_{00}$ is defined using the minimum-norm completion, it obeys 
\begin{equation} 
    \kappa_{00}\!\left(\ket{V_N^n}\right) \leq 
    \frac{ \left\|n\ket{V_N^{n-1}}\right\|^2 }{ \left\|\ket{V_N^n}\right\|^2 }
    =
    \frac{n}{N+1-n}.
    \label{eq:kappatower00-bound} 
\end{equation} 
This is the cost of the canonical tower completion supplied by \eqref{eq:lift}. 
with equality whenever the tower completion is the minimum-norm uplift. At fixed rung~$n$, the right-hand side decays as $n/N$. Similarly, the dressing cost for the $11$ channel is bounded by
\begin{equation}
    \label{eq:kappatower11-bound} 
    \kappa_{11}\leq \frac{N-n}{n+1}.
\end{equation}
The tower therefore provides an exact and explicitly dressed uplift, in contrast to the lossless uplift of the single-matrix model.

We now turn to the fortuitous states, where no closed form is available and the fidelity is computed numerically. Let $\mathcal S\subset\Bps_N^{(a,b)}$ be a fortuitous subspace in a given charge sector $(a,b)$. We evaluate the minimum uplift fidelity over its normalized states and over the disorder realizations,
\begin{equation}
\mathcal U_T(\mathcal S)
=
\min_s\;
\min_{\substack{\alpha\in\mathcal S\\ \|\alpha\|=1}}
\mathcal F_T^{(s)}(\alpha),
\label{eq:mT}
\end{equation}
where $s$ labels the disorder realization. Thus $\mathcal{U}_T(\mathcal S)=1$
means that every state in $\mathcal S$ lifts exactly through channel~$T$
in every realization included in the calculation. No sampling of states
within $\mathcal S$ is involved.

\begin{figure}[tb]
\centering
\usefig[\textwidth]{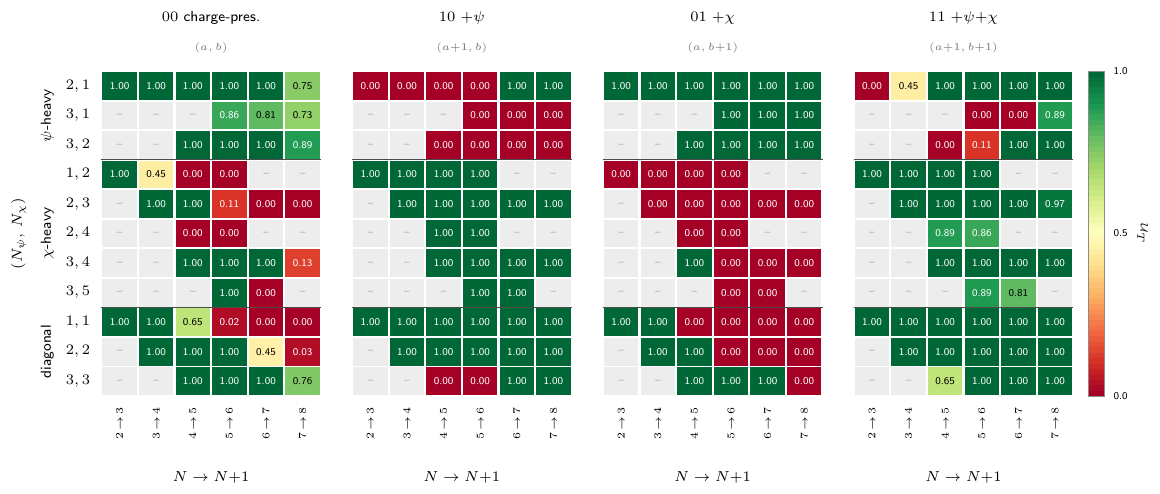}
\caption{Uplift fidelity $\mathcal{U}_T$ of \eqref{eq:mT} by charge sector (rows) and channel (panels). Green
($\mathcal{U}_T=1$) means every fortuitous state of that sector lifts exactly, in every realization; gray
dashes mark empty sectors. Each row is fully green in at least one panel. The minimum was taken over five disorder realizations.
}
\label{fig:four}
\end{figure}

Figure~\ref{fig:four} shows $\mathcal{U}_T$ for the fortuitous subspace of each starting charge sector and for each uplift channel. Reading across a row, every populated sector has at least one channel in which $\mathcal{U}_T=1$. Thus, within the range tested, every fortuitous subspace admits a channel through which all of its states lift exactly in all disorder realizations. Note that these state-resolved quantities depend on the chosen harmonic basis. However, the general trends are basis independent.

The pattern reflects the $\psi/\chi$ asymmetry of
$Q\sim\psi\psi\chi^\dagger$. The charge-preserving channel $00$ fails for the $\chi$-heavy sectors, whereas those sectors lift perfectly through the $+\psi$ channel. The $\psi$-heavy sectors behave in mirror image: they fail under $+\psi$ and  lift with $+\chi$.
In the cases studied, the successful single-mode channel shifts the charge toward the diagonal $N_\psi=N_\chi$. The $+\psi+\chi$ channel does not provide a
universal alternative: its minimum fidelity drops to zero on $(3,1)$ and $(3,2)$.

\begin{figure}[tb]
\centering
\usefig[\textwidth]{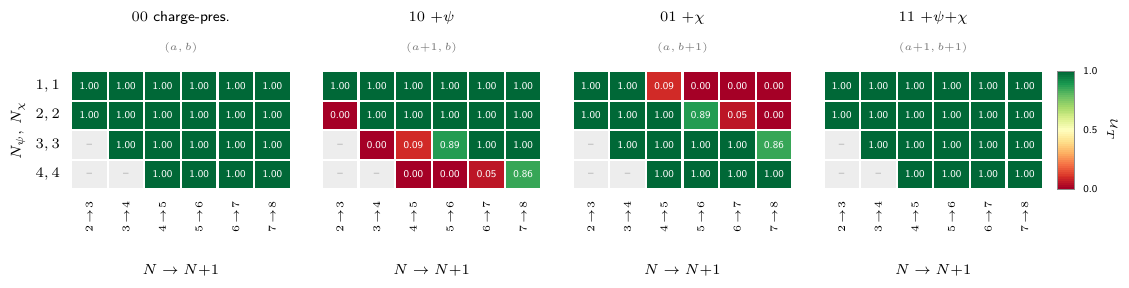}
\caption{Uplift fidelity $\mathcal{U}_T$ of the monotonous tower $V^n|\Omega\rangle$ on the diagonal
sectors, where the subspace is one-dimensional. The balanced channels $00$ and $11$ are green
everywhere, as forced by \eqref{eq:lift}; the single-mode channels $+\psi$ and $+\chi$ fail in
complementary, particle--hole-mirrored regimes. Contrast with Figure~\ref{fig:four}, where $00$ is
precisely the channel that fails.}
\label{fig:tower}
\end{figure}

Figure~\ref{fig:tower} applies the same diagnostic to the
monotonous tower.
At each diagonal charge $(n,n)$ the tower subspace is
one-dimensional, so $\mathcal{U}_T$ is
simply the fidelity of the single state $\ket{V_N^n}$. The fidelity of the $00$ and $11$ channels is one. Therefore, the $00$ and $11$ columns of Figure~\ref{fig:tower} are uniformly green. However, the two single-mode channels are not
protected. They break $\Npsi=\Nchi$ and they fail in complementary regimes: $+\psi$ is
obstructed when the tower is nearly full ($n\simeq N$), $+\chi$ in the dilute regime ($n\ll N$). The two panels
are related by particle--hole symmetry,
\begin{equation}
\mathcal F_{10}\!\left(\ket{V_N^n}\right)
=
\mathcal F_{01}\!\left(\ket{V_N^{N-n}}\right).
\end{equation}

We turn from whether a state lifts to how much it costs. On the exact-lift locus, we obtain a collection of dressing costs $\kappa_T$. The distribution contains bare directions at exactly $\kappa_T=0$, for which $\alpha\otimes\ket{T}$ is already BPS, together with a broad dressed population at $\kappa_T>0$. A minority of near-singular directions have costs extending several orders of magnitude above the bulk.
At each source sector, value of $N$, and channel, we show the $\kappa_T$ values as a point cloud, with the median marked by a horizontal bar. The vertical axis uses a symmetric-log coordinate, so that the bare directions at $\kappa_T=0$ remain visible on the baseline.

\begin{figure}[tbp]
\centering
\usefig[\textwidth]{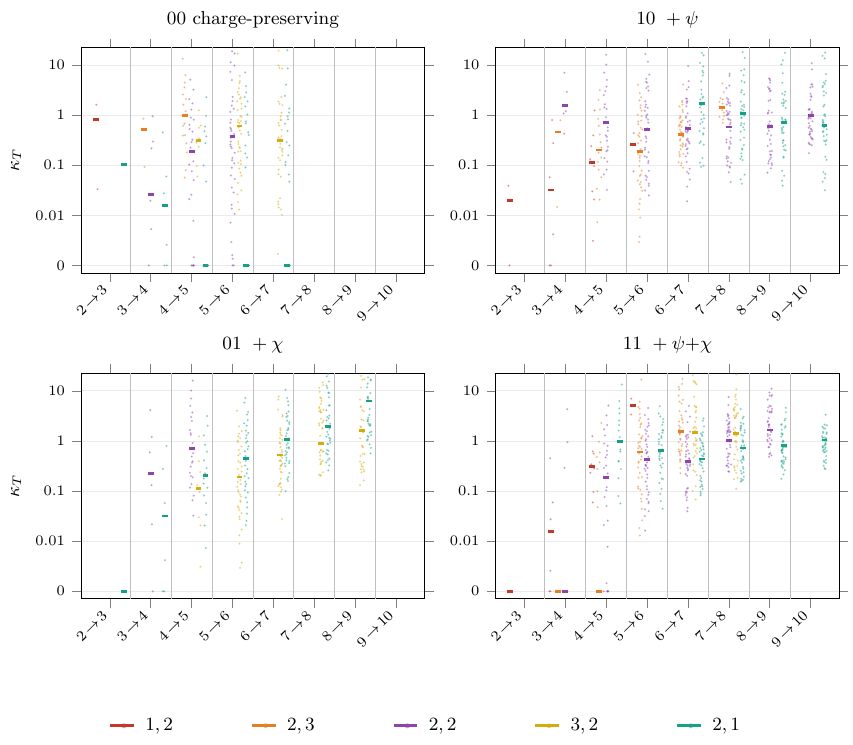}
\caption{Dressing cost of the fortuitous states, by uplift direction. At each step $N\to N+1$ the
five representative sectors $(N_\psi,N_\chi)$ are displaced horizontally and distinguished by color. For each we show 
the individual $\kappa_T$ values with the median indicated as
a bar. The sectors present differ between panels (cf.\ Fig.~\ref{fig:four}). The vertical axis uses a symmetric-log coordinate, allowing $\kappa_T=0$ to remain visible. The cost was computed with a single realization.}
\label{fig:dressfort}
\end{figure}

Figure~\ref{fig:dressfort} displays the fortuitous dressing costs by uplift direction for a representative set of five charge sectors: two $\chi$-heavy, two $\psi$-heavy, and one balanced. The sectors are displaced horizontally within each $N\to N+1$ region.
Only states with $\mathcal F_T=1$ in a given channel appear in that panel, so each carries a different,
$N$-dependent subset of sectors (cf.\ Figure~\ref{fig:four}). 
The clouds are broad and strongly channel-dependent. 
Furthermore, the median bars step up with $N$ in most channels. This means that the fortuitous states have no cheap protected
uplifts, and their typical dressing cost only grows as the theory is enlarged. This is in sharp
contrast to the tower, which does possess such a channel.

\begin{figure}[tbp]
\centering
\usefig[\textwidth]{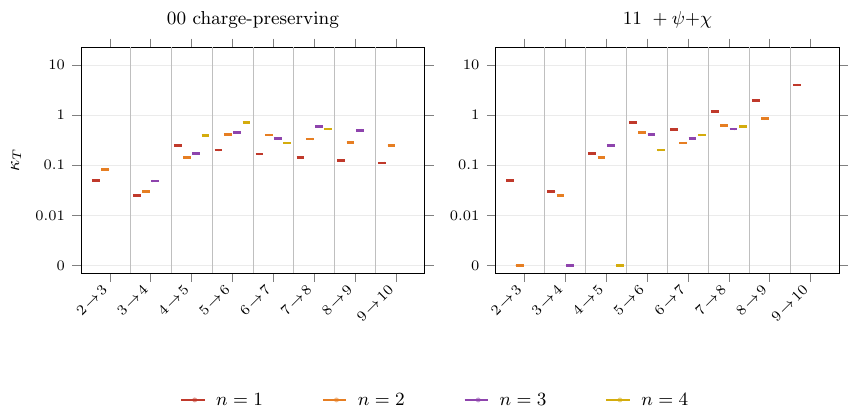}
\caption{Dressing cost of the monotonous tower in the two balance-preserving channels, $00$ (left) and $+\psi+\chi$ (right), for rungs $n=1,\ldots,4$. Unlike Figure~\ref{fig:dressfort}, the tower has a single state at each charge. For the $00$ channel, the cost asymptotically approaches $n/N$, while for the $11$ channel, it grows with $N$. The cost was computed with a single realization.}
\label{fig:dresstwo}
\end{figure}

Figure~\ref{fig:dresstwo} displays the dressing costs for the monotonous tower. The markers for the $00$ channel trace~\eqref{eq:kappatower00-bound}, while the markers for the $11$ channel trace~\eqref{eq:kappatower11-bound}. At fixed rung~$n$, the upper bound for the $00$ channel is suppressed as $n/N$ for $N\gg n$, while the second grows as $N/(n+1)$. In the $11$ channel, each rung has exactly zero cost at $N=n$ then steps up, while in the $00$ channel the low rungs settle onto the decaying $n/(N+1-n)$. Since $\kappa_T$ is defined by the minimum-norm uplift $D_T^{+}\alpha$, a cheaper completion may exist, so the measured value sits at or below
\eqref{eq:kappatower00-bound} and rises onto it only once $N\gtrsim n+3$.

\begin{figure}[tbp]
\centering 
\resizebox{0.9\textwidth}{!}{\definecolor{mc0}{RGB}{192,57,43}%
\definecolor{mc1}{RGB}{230,126,34}%
\definecolor{mc2}{RGB}{142,68,173}%
\definecolor{mc3}{RGB}{212,172,13}%
\definecolor{mc4}{RGB}{22,160,133}%
\definecolor{mc5}{RGB}{36,113,163}%
\definecolor{mc6}{RGB}{127,140,141}%
\begin{tikzpicture}
\begin{axis}[
  width=10cm, height=6.6cm,
  xmin=4.6, xmax=10.4, xtick={5,6,7,8,9,10},
  xlabel={$N$}, ylabel={$\displaystyle g_N$},
  ymode=log, ymin=0.08, ymax=3.6,
  ytick={0.1,0.3,1,3}, yticklabels={$0.1$,$0.3$,$1$,$3$},
  minor ytick={0.2,0.4,0.5,0.6,0.7,0.8,0.9,2},
  grid=both, major grid style={black!12}, minor grid style={black!4},
  tick align=outside, tick pos=left,
  every axis plot/.append style={line join=round},
  legend columns=1, legend cell align=left,
  legend style={
    at={(1.02,0.5)}, anchor=west, draw=none,
    fill=white, font=\footnotesize, column sep=1ex, /tikz/every even column/.append style={column sep=2ex},
  },
]
  \addplot[mc0,mark=*,mark size=1.5pt,thick] coordinates {(5,0.6373)(6,0.7204)(7,2.9050)};
  \addlegendentry{fortuitous $(2,1)$}
  \addplot[mc1,mark=*,mark size=1.5pt,thick] coordinates {(5,0.2554)(6,0.6294)(7,2.4590)};
  \addlegendentry{fortuitous $(1,2)$}
  \addplot[mc2,mark=*,mark size=1.5pt,thick] coordinates {(5,0.7497)(6,1.4370)(7,2.9120)};
  \addlegendentry{fortuitous $(2,2)$}
  \addplot[mc3,mark=*,mark size=1.5pt,thick] coordinates {(5,0.5556)(6,1.4040)(7,2.9590)};
  \addlegendentry{fortuitous $(3,2)$}
  \addplot[mc4,mark=*,mark size=1.5pt,thick] coordinates {(5,0.5413)(6,1.3850)(7,2.9140)};
  \addlegendentry{fortuitous $(2,3)$}
  \addplot[mc5,mark=square*,mark size=1.4pt,thick,densely dashed] coordinates {(5,0.2000)(6,0.1637)(7,0.1268)(8,0.0987)(9,0.1100)(10,0.1000)};
  \addlegendentry{tower $n=1$}
  \addplot[mc6,mark=triangle*,mark size=1.7pt,thick,densely dashed] coordinates {(5,0.4152)(6,0.2789)(7,0.3334)(8,0.2857)(9,0.2500)(10,0.2222)};
  \addlegendentry{tower $n=2$}
\end{axis}
\end{tikzpicture}}
\caption{Median greedy dressing cost defined in~\eqref{eq:marginal}, for states followed from $N_0=5$. Solid lines show the five fortuitous sectors,
greedily walked. The cost climbs steeply, and the walks are exactly continuable only through
$N=7$ here. Dashed lines show the tower rungs $n=1,2$, declining steadily out to $N=10$.}
\label{fig:marginal}
\end{figure}
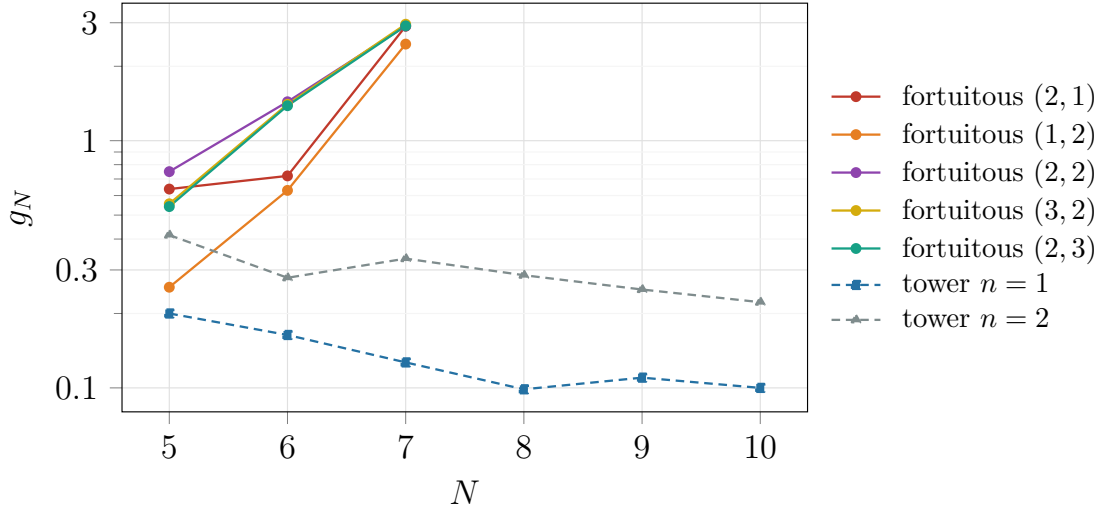

The preceding figures showed dressing costs for the BPS space in different charge sectors, but did not follow individual states as they are uplifted. To follow their evolution along successive uplifts, we track them individually across $N$. For a state $\alpha$, define the greedy dressing cost
\begin{equation}
g_N(\alpha)=\min_{T:\ \alpha\in\im D_T}\kappa_T(\alpha),
\label{eq:marginal}
\end{equation}
the cheapest way to carry $\alpha$ up one level over the directions in which it lifts exactly. We
seed with the fortuitous states of a sector at $N_0$ and walk each greedily, landing on the
min-norm uplift $D_{T^\star}^{+}\alpha$ and aggregating the median of $g_N$
over the states still alive at each $N$. For the tower we evaluate $g_N$ directly on the canonical
family $V_N^{\,n}|\Omega\rangle$, defined at every $N$ by \eqref{eq:lift}. Figure~\ref{fig:marginal}
shows the result, seeded at $N_0=5$: the fortuitous $g_N$ rises steeply, while the low tower
rungs decrease. The protected $00$ channel makes uplift marginally
cheaper as $N$ grows, exactly like $\kappa_{00}=n/(N{+}1{-}n)$. Thus the greedy dressing cost separates the two families in a way the sector averages do not: over the range tested, it increases along the surviving fortuitous walks but decreases along the protected tower.

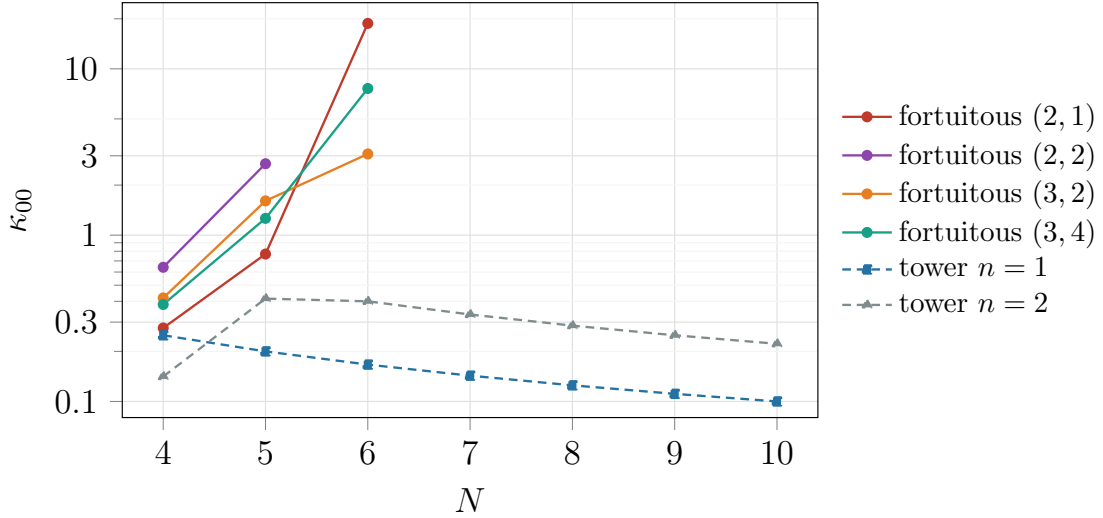
\begin{figure}[tbp]
\centering 
\resizebox{0.9\textwidth}{!}{\definecolor{mc0}{RGB}{192,57,43}%
\definecolor{mc1}{RGB}{142,68,173}%
\definecolor{mc2}{RGB}{230,126,34}%
\definecolor{mc3}{RGB}{22,160,133}%
\definecolor{mc4}{RGB}{36,113,163}%
\definecolor{mc5}{RGB}{127,140,141}%
\begin{tikzpicture}
\begin{axis}[
  width=10cm, height=6.6cm,
  xmin=3.6, xmax=10.4, xtick={4,5,6,7,8,9,10},
  xlabel={$N$}, ylabel={$\kappa_{00}$},
  ymode=log, ymin=0.08, ymax=25,
  ytick={0.1,0.3,1,3,10}, yticklabels={$0.1$,$0.3$,$1$,$3$,$10$},
  minor ytick={0.2,0.4,0.5,0.6,0.7,0.8,0.9,2,5,20},
  grid=both, major grid style={black!12}, minor grid style={black!4},
  tick align=outside, tick pos=left,
  every axis plot/.append style={line join=round},
  legend columns=1, legend cell align=left,
  legend style={
    at={(1.02,0.5)}, anchor=west, draw=none,
    fill=white, font=\footnotesize,
  },
]
  \addplot[mc0,mark=*,mark size=1.5pt,thick] coordinates {(4,0.2765)(5,0.7698)(6,18.7586)};
  \addlegendentry{fortuitous $(2,1)$}
  \addplot[mc1,mark=*,mark size=1.5pt,thick] coordinates {(4,0.6409)(5,2.6878)};
  \addlegendentry{fortuitous $(2,2)$}
  \addplot[mc2,mark=*,mark size=1.5pt,thick] coordinates {(4,0.4193)(5,1.6069)(6,3.0775)};
  \addlegendentry{fortuitous $(3,2)$}
  \addplot[mc3,mark=*,mark size=1.5pt,thick] coordinates {(4,0.3829)(5,1.2615)(6,7.6208)};
  \addlegendentry{fortuitous $(3,4)$}
  \addplot[mc4,mark=square*,mark size=1.4pt,thick,densely dashed] coordinates {(4,0.2500)(5,0.2000)(6,0.1666)(7,0.1429)(8,0.1250)(9,0.1111)(10,0.1000)};
  \addlegendentry{tower $n=1$}
  \addplot[mc5,mark=triangle*,mark size=1.7pt,thick,densely dashed] coordinates {(4,0.1412)(5,0.4153)(6,0.4000)(7,0.3333)(8,0.2857)(9,0.2500)(10,0.2222)};
  \addlegendentry{tower $n=2$}
\end{axis}
\end{tikzpicture}}
\caption{State-based dressing cost for walks restricted to the charge-preserving $00$ channel. Solid lines show the four fortuitous sectors that lift in $00$ for more than one step. Their
cost increases steeply and the walks survive only through $N=6$. Dashed lines show the tower rungs $n=1,2$, with $n=1$
tracing \eqref{eq:kappatower00-bound} exactly.}
\label{fig:marginal00}
\end{figure}

The minimization in~\eqref{eq:marginal} allows each state to choose its cheapest uplift channel, so a walk
may hop between directions from one step to the next. To remove this freedom, we can restrict the walk to the single
charge-preserving $00$ channel, which is the tower's protected direction. This places the two families on equal footing, at the price that a fortuitous sector need not lift in $00$ at every step and may drop out.
Figure~\ref{fig:marginal00} shows these channel-restricted walks from $N=4$ to $10$. Over the range tested, the surviving fortuitous walks become more expensive, whereas the low tower rungs become cheaper to uplift through the protected channel. The contrast observed after minimizing over channels is therefore also present within the charge-preserving channel alone.

\section{Metric fragility in the one-flavor $\mathcal{N}=2$ SYK model}
\label{sec:generic}

A BPS class may be uplifted indefinitely along a walk confined to the fortuity window. Physical states, however, are not just classes and this section asks whether they survive the uplift in one-flavor $\mathcal{N}=2$ SYK.  

Take a BPS state $\alpha$ of the size-$N$ theory and add one fermion. Is there
a genuine BPS state of the enlarged theory whose designated occupation
component is exactly $\alpha$? The decoder $D_T$ of Section~\ref{sec:rayleigh}
reads off that component, so we are asking whether $\alpha$ lies in its image.
These states form the exact-lift locus $\mathcal A^p_{N;T}$ defined in \eqref{eq:exact-lift-locus}, and they are precisely the states with $\cF_T=1$. Its codimension in the BPS space is given by 
\begin{equation} 
   \label{eq:codima}
    \operatorname{codim}_{\mathcal B_N^p}\mathcal A^p_{N;T} := \dim\mathcal B_N^p-\dim\mathcal A^p_{N;T}, 
\end{equation} 
which counts the number of independent BPS directions without an exact uplift through channel~$T$.

Throughout this section we take $q=3$ with generic couplings. For odd $N=2m+1$ the two
middle charge sectors $p=m$ and $p=m+1$ both hold $\dim\mathcal B_N^p=3^{m}$
states, at levels $\ell=2p-N=\mp1$.

\subsection{The constrained ambient spaces}

Naively, exact uplift should almost never work. The BPS space contains roughly $3^{N/2}$ states, whereas the ambient charge sector contains roughly $2^N$ states; two subspaces of such disparate dimensions, placed in generic position, would have little reason to intersect. It is therefore surprising that, in the odd-$N$ middle sector, the channel directed toward the center of the fortuity window contains the entire BPS space through $N=9$.

The surprise dissolves once one notices that both subspaces are confined by
the supersymmetry conditions before any comparison is made. Writing a state
of the enlarged theory as $\Psi=(\alpha,\beta)$, the requirements
$Q_{N+1}\Psi=Q_{N+1}^\dagger\Psi=0$ force $Q_N\alpha=0$ and
$Q_N^\dagger\beta=0$ on the two blocks separately. Hence, for the down and up
channels, respectively,
\begin{equation}\label{eq:ambient}
  \im D_{\downarrow}
  \subseteq
  \mathcal R_{N;\downarrow}^p
  :=
  \ker Q_N\big|_{\HH_N^p},
  \qquad
  \im D_{\uparrow}
  \subseteq
  \mathcal R_{N;\uparrow}^p
  :=
  \ker Q_N^\dagger\big|_{\HH_N^p}.
\end{equation}
The dimensions of these constrained ambient spaces are
\begin{equation}\label{eq:room-dimensions}
  \dim\mathcal R_{N;\downarrow}^p
  =
  \binom Np-\rk M_p,
  \qquad
  \dim\mathcal R_{N;\uparrow}^p
  =
  \dim\mathcal B_N^p+\rk M_p,
\end{equation}
where
$M_p:\HH_N^p\to\HH_N^{p+q}$ is the matrix of $Q_N$ acting from charge $p$.
The BPS space lies in both constrained rooms. Generic position counting should therefore be performed
inside these smaller spaces. If $\mathcal B_N^p$ and $\im D_T$ are in generic
position inside $\mathcal R_{N;T}^p$, their predicted intersection dimension
is
\begin{equation}\label{eq:generic-intersection}
  a_{\mathrm{gen}}
  :=
  \max\left\{
    0,\,
    \dim\mathcal B_N^p+\rk D_T-\dim\mathcal R_{N;T}^p
  \right\}.
\end{equation}

Table~\ref{tab:exact-lift} shows that inside the
constrained rooms the observed intersections agree with the generic prediction
in almost all cases. For instance,
$(N,p)=(11,5)$ inside $\ker Q_{11}^\dagger$ one predicts $243+406-407=242$ and
observes exactly that.
Exceptions first occur at $(9,3)$ and its mirror $(9,6)$, both sitting on the window edge
$|\ell|=q$ where the exceptional states live, and the surplus is a single
dimension. 
Every other sector in the restricted fortuity window matches the generic prediction exactly for $N \le 13$.

The channel that reaches the middle of the fortuity window ($\ell' = \ell\pm 1 =0$)
keeps the whole BPS space up to $N=9$, and first loses a single direction at
$N=11$.
The obstruction is mild but nonzero: in the odd-$N$ middle sector at $N=11$, a generic BPS state has $\cF_\uparrow<1$ in the middle-directed up channel.

The obstruction can be expressed directly in terms of the block supersymmetry conditions. Work in the up channel, so that
$\alpha\in\mathcal B_N^p$ is the occupied component of
$\Psi=(\alpha_\varnothing,\alpha)$ at charge $p+1$, and the unknown is the
correction $\alpha_\varnothing$. We recall from Section~\ref{sec:hilbert-grow} that $C'$ denotes the part of the supercharge built from couplings involving the added mode. For $q=3$, it raises the fermion number of the original modes by two, while $C'^\dagger$ lowers it by two. Using $Q_N\alpha=Q_N^\dagger\alpha=0$, the
block forms \eqref{eq:block-matrix} reduce supersymmetry of $\Psi$ to three
conditions on that correction,
\begin{equation}\label{eq:three-conditions}
    Q_N\alpha_\varnothing=0,
  \qquad
   C'\alpha_\varnothing=0,
  \qquad
   Q_N^\dagger\alpha_\varnothing=-C'^\dagger\alpha .
\end{equation}
The third condition has the clearest physical interpretation: the fresh couplings acting on $\alpha$ produce the source $C'^\dagger\alpha$, which must be canceled by $Q_N^\dagger\alpha_\varnothing$. It also gives a subspace that uplifts at no cost.
If $C'^\dagger\alpha=0$ the fresh couplings inject nothing and the bare embedding is already BPS. Hence,
\begin{equation}\label{eq:free-subspace}
  \ker C'^\dagger\big|_{\mathcal B_N^{p}}\subseteq\mathcal A^p_{N;\uparrow},
\end{equation}
and counting puts its dimension at least
$\dim\mathcal B_N^{p}-\binom{N}{p-q+1}$. These are the SYK analog of the
staircase channel of Section~\ref{sec:chen-model}: rigid, and isometric with zero dressing cost.

\begin{table}[t]
\centering\small
\begin{tabular}{cccc|ccccc|ccccc}

\multicolumn{4}{c}{} &
\multicolumn{5}{c}{down channel} &
\multicolumn{5}{c}{up channel}\\

\multicolumn{1}{c}{$N$} &
\multicolumn{1}{c}{$p$} &
\multicolumn{1}{c}{$\ell$} &
\multicolumn{1}{c}{$\dim\mathcal B_N^p$} &
\multicolumn{1}{c}{$\dim\mathcal R$} &
\multicolumn{1}{c}{$\rk D$} &
\multicolumn{1}{c}{\hspace{-0.2cm}$a_{\rm gen}$} &
\multicolumn{1}{c}{\hspace{-0.2cm}$\dim\mathcal A$} &
\multicolumn{1}{c}{\hspace{-0.1cm}codim} &
\multicolumn{1}{c}{$\dim\mathcal R$} &
\multicolumn{1}{c}{$\rk D$} &
\multicolumn{1}{c}{\hspace{-0.2cm}$a_{\rm gen}$} &
\multicolumn{1}{c}{\hspace{-0.2cm}$\dim\mathcal A$} &
\multicolumn{1}{c}{\hspace{-0.1cm}codim}\\
\midrule
\rowcolor{yellow!25}
$5$  & $1$ & $-3$ & $1$
     & $1$    & $0$    & $0$   & $0$   & $1$
     & $5$    & $5$    & $1$   & $1$   & $0$\\
$5$  & $2$ & $-1$ & $9$
     & $9$    & $8$    & $8$   & $8$   & $1$
     & $10$   & $10$   & $9$   & $9$   & $0$\\
$6$  & $2$ & $-2$ & $9$
     & $9$    & $0$    & $0$   & $0$   & $9$
     & $15$   & $15$   & $9$   & $9$   & $0$\\
$6$  & $3$ & $0$  & $18$
     & $19$   & $19$   & $18$  & $18$  & $0$
     & $19$   & $19$   & $18$  & $18$  & $0$\\
$7$  & $3$ & $-1$ & $27$
     & $28$   & $27$   & $26$  & $26$  & $1$
     & $34$   & $34$   & $27$  & $27$  & $0$\\
$8$  & $3$ & $-2$ & $27$
     & $28$   & $3$    & $2$   & $2$   & $25$
     & $55$   & $55$   & $27$  & $27$  & $0$\\
$8$  & $4$ & $0$  & $54$
     & $62$   & $62$   & $54$  & $54$  & $0$
     & $62$   & $62$   & $54$  & $54$  & $0$\\
     \rowcolor{yellow!25}
$9$  & $3$ & $-3$ & $3$
     & $4$    & $0$    & $0$   & $0$   & $3$
     & $83$   & $81$   & $1$   & \cellcolor{orange!40} 2 & $1$\\
$9$  & $4$ & $-1$ & $81$
     & $90$   & $81$   & $72$  & $72$  & $9$
     & $117$  & $117$  & $81$  & $81$  & $0$\\
$10$ & $4$ & $-2$ & $81$
     & $91$   & $0$    & $0$   & $0$   & $81$
     & $200$  & $199$  & $80$  & $80$  & $1$\\
$10$ & $5$ & $0$  & $162$
     & $207$  & $207$  & $162$ & $162$ & $0$
     & $207$  & $207$  & $162$ & $162$ & $0$\\
$11$ & $5$ & $-1$ & $243$
     & $298$  & $243$  & $188$ & $188$ & $55$
     & $407$  & $406$  & $242$ & $242$ & $1$\\
$12$ & $5$ & $-2$ & $243$
     & $309$  & $1$    & $0$   & $0$   & $243$
     & $726$  & $713$  & $230$ & $230$ & $13$\\
$12$ & $6$ & $0$  & $486$
     & $705$  & $704$  & $485$ & $485$ & $1$
     & $705$  & $704$  & $485$ & $485$ & $1$\\
     \rowcolor{yellow!25}
$13$ & $5$ & $-3$ & $1$
     & $79$   & $0$    & $0$   & $0$   & $1$
     & $1209$ & $729$  & $0$   & $0$   & $1$\\
$13$ & $6$ & $-1$ & $729$
     & $1014$ & $729$  & $444$ & $444$ & $285$
     & $1431$ & $1417$ & $715$ & $715$ & $14$\\
\hline
\end{tabular}
\caption{\label{tab:exact-lift}%
Exact-lift loci for every sector with  $\ell\equiv 2p-N\le0$. Here $\dim\mathcal R$ is the dimension of the constrained ambient space,  $a_{\rm gen}$ is the exact-lift-locus dimension predicted when $\mathcal B_N^p$ and $\im D_T$ are in generic position inside $\mathcal R_{N;T}^p$, and ``codim'' is the codimension defined in~\eqref{eq:codima}. Sectors with $\ell>0$
follow by particle--hole conjugation $p\mapsto N-p$.
The observed dimensions agree with the generic position prediction ($\dim\mathcal A=a_{\rm gen}$) everywhere except at $(N,p)=(9,3)$, where the intersection exceeds the prediction by one. 
}
\end{table}

\subsection{Asymptotic obstruction to exact uplift}
\label{sec:permanent}

We now specialize the generic position prediction
\eqref{eq:generic-intersection} to the up channel. The full BPS space is
contained in the predicted exact-lift locus when $D_\uparrow$ fills the
constrained room $\mathcal R_{N;\uparrow}^p$. BPS directions are lost when
\[
  \rk D_\uparrow
  <
  \dim\mathcal R_{N;\uparrow}^p,
\]
and the predicted exact-lift locus vanishes when
\[
  \dim\mathcal R_{N;\uparrow}^p-\rk D_\uparrow
  \geq
  \dim\mathcal B_N^p.
\]

The up-channel room has dimension given in \eqref{eq:room-dimensions},
while $D_\uparrow$ is defined on $\mathcal B_{N+1}^{p+1}$. Assuming that
the relevant connecting maps vanish, its domain has dimension
\[
  \dim\mathcal B_{N+1}^{p+1}
  =
  \dim\mathcal B_N^{p+1}+\dim\mathcal B_N^p.
\]
Since the rank of a linear map cannot exceed the dimension of its domain,
\begin{equation}\label{eq:room-deficit}
  \dim\mathcal R_{N;\uparrow}^p-\rk D_\uparrow
  \geq
  \rk M_p-\dim\mathcal B_N^{p+1}.
\end{equation}
This bound gives sufficient conditions both for the loss of some BPS
directions and for the complete disappearance of the predicted exact-lift
locus.

Take $N=2m+1$ at $\ell=-1$. We have
$\dim\mathcal B_N^{m}=\dim\mathcal B_N^{m+1}=3^m$, and
Conjecture~\ref{conj:rank} gives $\rk M_m=\Dtil_{m-2}$, where $\widetilde D_m$
is the index-truncated dimension introduced in \eqref{eq:Dtil}.
Hence, the right-hand side
of \eqref{eq:room-deficit} is $\Dtil_{m-2}-3^m$. The two thresholds are
therefore
\begin{equation}\label{eq:sufficient}
  3^m<\Dtil_{m-2}
  \qquad\text{and}\qquad
  2\cdot 3^m\le\Dtil_{m-2},
\end{equation}
the first obstructing full uplift and the second forcing the exact-lift
locus to vanish.

The obstruction follows from the different asymptotic growth rates. The BPS space grows as
$3^{N/2}$ while the relevant supercharge rank is of order $2^N$, so
$\Dtil_{m-2}/3^m$ grows as $(4/3)^m$ up to a power of $m$ and eventually
exceeds any fixed value. Evaluating both sides, the first condition is first satisfied at $m=7$ and the second at $m=9$. Hence, subject to
Conjecture~\ref{conj:rank}, the vanishing of the relevant connecting maps, and
generic position, we expect 
\begin{equation}\label{eq:permanent}
  \operatorname{codim}_{\mathcal B_N^m}\mathcal A^m_{N;\uparrow}>0
  \quad\text{for odd }N\ge15,
  \qquad
  \mathcal A^m_{N;\uparrow}=\{0\}
  \quad\text{for odd }N\ge19 .
\end{equation}

Nothing dynamical happens at either threshold: one exponential simply overtakes
another. For generic couplings the middle-directed up channel eventually
carries no nonzero harmonic representative exactly, which does not exclude
exact uplift through another channel or at specially tuned couplings.
In this sense metric fragility in the generic model has an arithmetic origin.
The conditions are sufficient but not necessary: a direction is lost at $N=11$, well below the first
threshold.

\begin{figure}[t]
  \centering
  \includegraphics[width=\textwidth]{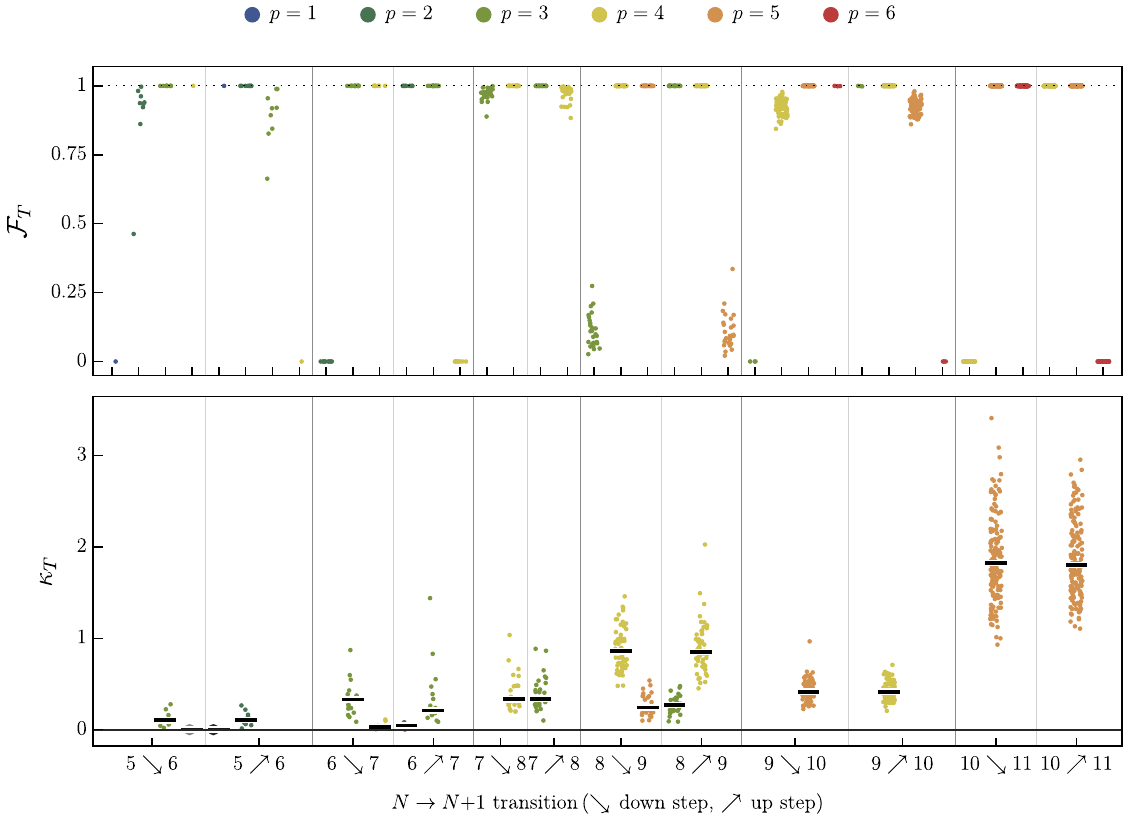}
  \caption{\label{fig:state-diagnostics}
  Fidelity and dressing cost for one-step transitions
  $N\to N+1$ in the one-flavor $\mathcal N=2$ SYK model. Colors label the
  initial fermion number $p$. The symbols $\searrow$ and $\nearrow$ denote the down and up
  channels, respectively. \emph{Top:} uplift fidelity $\cF_T$ for every state in the
  chosen orthonormal harmonic basis. \emph{Bottom:} exact dressing costs $\kappa_T$ in sectors
  for which every displayed basis state has unit fidelity. Horizontal bars
  indicate sector medians.
  Columns without points are sectors where the exact uplift fails.}
\end{figure}

\subsection{Cost of the approximate uplift}
\label{sec:fidelity-cost}

We now evaluate the decoder diagnostics state by state. In each charge sector $\mathcal B_N^p$, we choose an orthonormal harmonic basis $\{\alpha_i\}$ and evaluate $\cF_T(\alpha_i)$ in both up and down channels, together with $\kappa_T(\alpha_i)$ whenever $\cF_T(\alpha_i)=1$. These state-resolved values depend on the chosen basis, whereas $\dim\mathcal A^p_{N;T}$ and the decoder singular-value spectrum are basis independent.

Figure~\ref{fig:state-diagnostics} summarizes the one-step up- and down-channel results for a single realization of the couplings. The upper panel shows the fidelity of every basis state. A sector lies entirely on $\cF_T=1$ precisely when its full BPS space is contained in $\im D_T$. Once this containment fails, at least one basis direction has $\cF_T<1$, and the harmonic states in that sector cannot all be uplifted exactly through the chosen channel.

The lower panel shows $\kappa_T$ only for sectors in which every displayed basis state lifts exactly. Each point gives the cost of the minimum-norm exact uplift, and the horizontal bars mark sector medians. Points at $\kappa_T=0$ correspond to bare embeddings that are already harmonic. A blank column means that the corresponding sector is not wholly contained in the exact-lift locus. In particular, the first failure in the middle-directed channel occurs for the transition $N=11\to12$, where one BPS direction loses unit fidelity. For states with $\cF_T<1$, one could instead study the dressing cost of the state $P_{\im D_T}\alpha$, but this would measure the uplift of the visible projection rather than of the original state.

The displayed exact costs tend to increase over the accessible transitions.
Determining whether an appropriately generalized dressing cost grows without bound or remains of order one is left for future work. In either case, the available data show no indication of the asymptotically vanishing cost found for the monotonous tower in the two-flavor model.

\begin{figure}[t]
  \centering
  \includegraphics[width=0.6\textwidth]{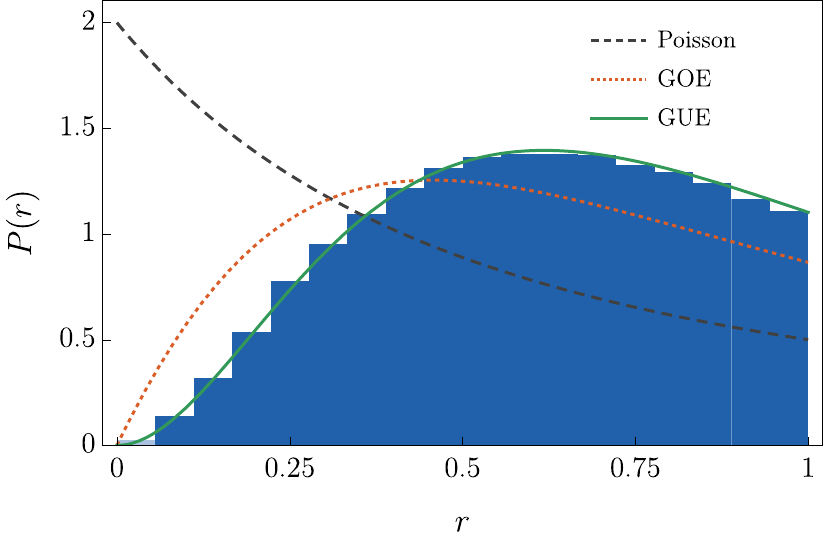}
  \caption{\label{fig:decoder-chaos}
  Adjacent-gap statistics of the down-channel decoder spectrum in
  the central sector $N'=12$, $P=6$, pooled over 500
  disorder realizations. Eigenvalues at the endpoints
  $\lambda=0,1$ are excluded. The probability density $P(r)$ is compared with
  the Poisson (dashed), GOE (red dotted), and GUE (green solid) ratio distributions.}
\end{figure}

\subsection{BPS chaos of the decoder spectrum}
\label{sec:decoder-chaos}

The decoder also defines a natural operator acting entirely within the enlarged
BPS space,
\begin{equation}
  \mathcal O_T
  :=
  D_T^\dagger D_T
  =
  P_{\mathcal B_{N'}^P}\,
  \Pi_T^\dagger\Pi_T\,
  P_{\mathcal B_{N'}^P}.
  \label{eq:decoder-lmrs}
\end{equation}
This is an operator of Lin--Maldacena--Rozenberg--Shan type: a simple occupation-pattern projector compressed to a highly degenerate BPS subspace \cite{LMRS,Lin:2022rzw,ChenLinShenker}. Projected operators of this kind have also been used to resolve a chaos--integrability crossover within the BPS subspace of an interpolating $\mathcal N=2$ SYK model \cite{Miyahara:2026iso}. Our perspective is different. Here $\mathcal O_T$ is fixed canonically by an inter-size decoder rather than chosen only as a probe of chaos. Its nonzero eigenvalues are the squared singular values $\lambda_j=\sigma_j^2$ of the decoder and coincide with the nonzero spectrum of $K_T=D_TD_T^\dagger$. These eigenvalues determine the possible visibility and dressing scales across system sizes, while their level correlations provide a simultaneous probe of chaos within the enlarged BPS space.

For one added fermion, the two channel operators are
\begin{equation}
  \mathcal O_{\downarrow}
  =
  P_{\mathcal B_{N'}^P}(1-n_{N'})P_{\mathcal B_{N'}^P},
  \qquad
  \mathcal O_{\uparrow}
  =
  P_{\mathcal B_{N'}^P}n_{N'}P_{\mathcal B_{N'}^P},
  \label{eq:decoder-channel-operators}
\end{equation}
and satisfy $
  \mathcal O_{\downarrow}+\mathcal O_{\uparrow}
  =
  \mathbf 1_{\mathcal B_{N'}^P}$. 
The two operators therefore share an eigenbasis, with complementary eigenvalues. The two channels consequently have identical adjacent-gap statistics, so it is sufficient to analyze one of them.

We study the down-channel spectrum after removing the eigenvalues
at $\lambda=0$ and $\lambda=1$. Ordering the remaining eigenvalues as
$\lambda_1<\lambda_2<\cdots$, we define
\begin{equation}
  s_i:=\lambda_{i+1}-\lambda_i,
  \qquad
  r_i:=
  \frac{\min(s_i,s_{i+1})}{\max(s_i,s_{i+1})}
  \in[0,1].
  \label{eq:decoder-gap-ratio}
\end{equation}
The adjacent-gap ratio requires no unfolding of the spectrum. Its standard
mean values for different ensemble types are \cite{ataspaper}
\begin{equation}
  \langle r\rangle_{\mathrm{Poisson}}=0.3863,
  \qquad
  \langle r\rangle_{\mathrm{GOE}}=0.5307,
  \qquad
  \langle r\rangle_{\mathrm{GUE}}=0.5996.
  \label{eq:gap-ratio-benchmarks}
\end{equation}

For the central sector of the enlarged theory at $N'=12$ and $P=6$, we
computed the decoder spectrum for 500
coupling realizations. After removing the endpoint modes, we find
\begin{equation}
  \langle r\rangle
  =
  0.599\pm0.001,
  \label{eq:decoder-mean-r}
\end{equation}
where the uncertainty is the standard error over the disorder realizations. The full adjacent-gap distribution likewise strongly favors the GUE benchmark, as shown in Figure~\ref{fig:decoder-chaos}.

The nontrivial decoder spectrum in this sector therefore exhibits level statistics consistent with the unitary random matrix class. Unlike a projected operator introduced solely to diagnose BPS chaos, however, $\mathcal O_T$ is determined by the uplift problem itself. Its eigenvalues determine metric uplift between system sizes, while correlations among the same eigenvalues diagnose chaos within the enlarged BPS space. The decoder thus ties two a priori distinct questions, inter-size reconstruction and BPS chaos, to different spectral observables of a single operator. A systematic study of other charges and larger sizes would be needed to determine the finite-size crossover and Thouless scale.

\section{Discussion}
\label{sec:discussion}

 \paragraph{Summary.}
In this paper, we have studied fortuity in SYK models and separated it into two layers: 
the algebraic uplift of BPS classes and the metric continuation of their harmonic
representatives. Whenever neighboring cohomologies are related by the long
exact sequence, fortuitous states are substantially more
robust than a fixed-charge analysis would suggest. Inside the fortuity
window, at least one of the two connecting homomorphisms vanishes, so every
nonzero BPS class admits a one-step uplift. Iterating these steps,
each class can be uplifted to arbitrarily large $N$.

At the level of normalized states, however, cohomological uplift is
not sufficient. The decoder of Section~\ref{sec:rayleigh} supplies the missing
metric information. For a channel $T$, the uplift fidelity
$\cF_T(\alpha)$ determines the fraction of a state visible to the
decoder and equals one precisely when the enlarged theory contains a BPS
state whose designated component is $\alpha$. When such an exact
uplift exists, the dressing cost $\kappa_T(\alpha)$ measures the additional
squared norm required by its minimum-norm realization upstairs. Together
with the bare fraction $f_{0,T}(\alpha)$, these diagnostics are moments of a single state-dependent decoder spectral measure.

The analytically controlled models illustrate several distinct forms of
metric continuation. In the single-matrix model, protected states admit a
bare isometric embedding and therefore uplift with unit fidelity and zero
dressing cost. In the protected two-flavor tower, exact continuation remains
possible but requires a symmetry-controlled dressing, whose cost can be
computed explicitly and decreases at fixed position in the tower.  
The generic one-flavor SYK model behaves differently: its exact-lift locus eventually vanishes. Thus BPS classes can remain algebraically continuable even when their harmonic representatives cease to admit exact uplift through that channel.

Finally, the decoder connects metric fragility with the
spectral structure of the BPS sector. The associated operator
\begin{equation}
  D_T^\dagger D_T
  =
  P_{\mathcal B_{N'}^P}\,
  \Pi_T^\dagger\Pi_T\,
  P_{\mathcal B_{N'}^P},
\end{equation}
is of Lin--Maldacena--Rozenberg--Shan type: it is a simple
occupation-pattern projector compressed to the enlarged BPS subspace~\cite{LMRS,Lin:2022rzw}. Its eigenvalues
control visibility and reconstruction cost, while correlations among them 
probe BPS chaos. In the generic one-flavor model, the
nontrivial decoder spectrum exhibits adjacent-gap statistics consistent with
the unitary random-matrix class. These results suggest that metric fragility provides a diagnostic of chaotic BPS dynamics across system sizes and complements other probes of BPS chaos, including projected-operator statistics \cite{LMRS,ChenLinShenker, Miyahara:2026iso}, non-Abelian Berry curvature under coupling deformations~\cite{Chen:2026vml}, and random matrix descriptions of fortuitous BPS sectors~\cite{Johnson:2026plw}.

 \medskip
 
These results raise several questions about the generality and physical interpretation of metric fragility.

\paragraph{Metric fragility as a signature of BPS chaos.} 
The contrast among the three models suggests a sharp conjecture. Protected or integrable BPS families admit rigid maps across $N$, realized either by an isometric bare embedding or by a symmetry-controlled dressing. In a chaotic BPS sector, by contrast, the harmonic subspace may rotate irregularly as the system is enlarged, making exact uplift  noncanonical, increasingly costly, and potentially obstructed. We therefore conjecture that metric fragility is a general signature of chaotic BPS dynamics. Testing this conjecture will require comparing decoder spectra across a broader range of charges, system sizes, and supersymmetric models.

\paragraph{Metric fragility in the two-flavor model.} The analysis of Section~\ref{sec:generic} could be extended to the symmetrized two-flavor model of Section~\ref{sec:twoflavor}. Its protected monotonous tower admits exact uplift through the balanced channels at every $N$, with a dressing cost that vanishes asymptotically. The fortuitous states lack this protection and are expected eventually to lose exact uplift, as in the generic one-flavor model. Our numerics reach only $N=8$, where every fortuitous sector still uplifts exactly through at least one channel, so an open problem is to determine whether and where metric fragility first appears in this model.

\paragraph{Fortuity and fragility beyond SYK.}
Nothing in the definition of the uplift fidelity is specific to SYK: it
requires only a family of theories related by adding fermionic modes, a
conserved $U(1)$ charge, and a BPS subspace in each theory. 
It would be very interesting to formulate the metric
refinement in the D1--D5 system, where fortuity has been exhibited~\cite{ChangLinZhang}, and in
ABJM and related settings~\cite{ABJMfortuity,HughesShigemori}.

\begin{table}[t]
\centering
\begin{tabular}{c|ccccc}
  $n\,\backslash\,m$ & $3$ & $4$ & $5$ & $6$ & $7$ \\ \hline
  $0$ & $1$ & $1$ & $1$ & $1$ & $1$ \\[2pt]
  $1$ & $\mathbf{2}$ & $0$ & $0$ & $0$ & $0$ \\[2pt]
  $2$ & $1$ & $2+\mathbf{20}$ & $2$ & $2$ & $2$ \\[2pt]
  $3$ & & $0$ & $\mathbf{106}$ & $0$ & $0$ \\[2pt]
  $4$ & & $1$ & $2$ & $3+\mathbf{711}$ & $3$ \\[2pt]
  $5$ & & & $0$ & $0$ & $\mathbf{4716}$
\end{tabular}
\caption{Dimensions $\dim H^n(\mathcal M_m)$ for the Abelian three-node cyclic quiver with $m$ arrows per edge. Bold entries denote the variable contribution to the middle cohomology, illustrating $R$-charge concentration, while the remaining nonzero entries are inherited from the ambient projective space.}
\label{tab:quiver-cohomology}
\end{table}

Another simple geometric system exhibiting $R$-charge concentration is quiver quantum mechanics~\cite{Denef:2002ru,Bena:2012hf,Manschot:2012rx,Anninos:2016szt}. For
the Abelian three-node cyclic quiver with $m$ arrows per edge and a generic
superpotential, the Higgs-branch moduli space $\cM_m$ is a complete
intersection in $\mathbb{CP}^{m-1}\times\mathbb{CP}^{m-1}$ of complex
dimension $m-2$. In every degree $n\neq m-2$, its cohomology is inherited from the projective ambient space and, at fixed degree, stabilizes as $m$ increases, with only polynomially growing dimension. The variable part of the cohomology is concentrated in the middle degree $n=m-2$ and corresponds to the pure-Higgs sector studied in~\cite{Bena:2012hf,Manschot:2012rx}. At fixed degree~$n$, variable classes therefore occur at exactly one value, $m=n+2$, giving a fortuity window of width one whose dimension grows rapidly with~$n$.

  The pattern is summarized in Table~\ref{tab:quiver-cohomology}, which displays
$\dim H^n(\cM_m)$, computed from the standard cohomology theory of complete
intersections, with the variable contributions in bold.
With the degree~$n$ playing the role of the $R$-charge, the table shows the same
structure as the SYK data of Figure~\ref{fig:bpsdatare}: monotonous (ambient)
classes persisting horizontally, and fortuitous (variable) classes
concentrated on a diagonal window of width one. This suggests a diagonal uplift of the variable classes, shifting the degree by one with each unit step in~$m$, as an analog of the walks in Section~\ref{sec:walks}. Constructing this uplift and developing its metric description remain interesting open problems.

\paragraph{Quantum error correction.} The decoder also admits a quantum information theoretic interpretation. Passing from the enlarged system at $N'$ back to the original system at $N$ amounts to removing  the additional fermionic modes. This defines a nonspatial erasure process, and metric fragility measures the recoverability of BPS information after this erasure. A natural question is whether there exist BPS subspaces on which the decoder singular values are approximately uniform, so that arbitrary superpositions uplift with approximately state-independent cost. This would test whether the uplift preserves not only individual BPS states but also their superpositions. It would be interesting to relate this form of protection under mode erasure to holographic quantum error correction.

\paragraph{Non-BPS states.} Although our analysis has focused on BPS states, the metric framework is not intrinsically cohomological. More generally, one may choose any physically distinguished target subspace at system size $N+1$, such as a spectral band of the Hamiltonian, and define the corresponding decoder by projecting onto that subspace. This may provide a systematic framework for studying the uplift of non-BPS states across $N$, including whether spectral bands can be identified between neighboring theories and whether their uplift becomes increasingly lossy or  expensive. We leave this extension for future investigation.

\paragraph{Gravitational interpretation.} Since monotonous states are naturally associated with perturbative multigraviton states, while fortuitous states are candidate black hole microstates, it is natural to ask whether metric fragility has a bulk interpretation. In particular, the contrast between protected and chaotic uplift may distinguish semiclassical states from strongly mixed black hole microstates. Making this connection precise would require a gravitational construction of the decoder, which we leave for future work.

\section*{Acknowledgments}

The authors are grateful to T.~Anous, M.~Buican, C.-M.~Chang, M.~Hanada,
M.~Hosseini, J.~Inglis, R.~Russo, K.~H.~Thong, and D.~Turton for interesting discussions.
DV is partially supported by the STFC Consolidated Grant
ST/X00063X/1 ``Amplitudes, strings \& duality.'' AI tools assisted with developing code, checking computations, and editing the manuscript. \\

\noindent \textbf{Data access statement.} The data used in this study are publicly available in the
ancillary files at \url{https://arxiv.org/abs/2608.12160}.  

\appendix
\section{Walk counting of the Hodge sectors}
\label{app:walks}

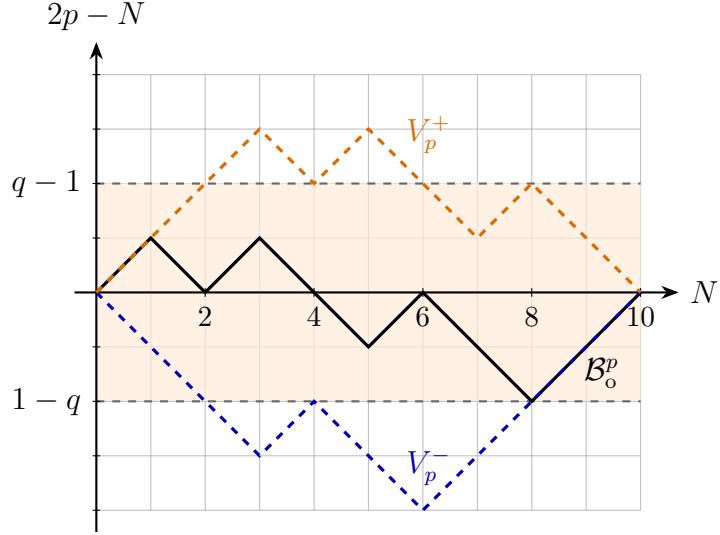
\begin{figure}[bt]
\centering
\begin{tikzpicture}[scale=0.72, >=Stealth]
\draw[gray!50, thin] (0,-4) grid (10,4);
\fill[orange!15, opacity=0.7] (0,-2) rectangle (10,2);
\draw[dashed, black!60, thick] (0, 2) -- (10, 2);
\draw[dashed, black!60, thick] (0,-2) -- (10,-2);
\draw[->, thick] (-0.4, 0) -- (10.7, 0) node[right, font=\normalsize]{$N$};
\draw[->, thick] (0, -4.4) -- (0, 4.6)  node[above, font=\normalsize]{$2p-N$};
\foreach \y in {-4,-3,-2,-1,1,2,3,4} { \draw (0.07,\y) -- (-0.07,\y); }
\node[left, font=\normalsize] at (-0.08,  2) {$q-1$};
\node[left, font=\normalsize] at (-0.08, -2) {$1-q$};
\foreach \n in {2,4,6,8,10} {
    \draw (\n, 0.08) -- (\n,-0.08) node[below, font=\small]{$\n$};
}
\draw[black, very thick]
    (0,0)--(1,1)--(2,0)--(3,1)--(4,0)--(5,-1)--(6,0)--(7,-1)--(8,-2)--(9,-1)--(10,0);
\node[below left, font=\normalsize, black] at (9.8, -1) {$\mathcal{B}^p_{\mathrm{o}}$};
\draw[blue!70!black, very thick, dashed]
    (0,0)--(1,-1)--(2,-2)--(3,-3)--(4,-2)--(5,-3)--(6,-4)--(7,-3)--(8,-2)--(9,-1)--(10,0);
\node[right, font=\normalsize, blue!70!black] at (5.5, -3.2) {$V^-_p$};
\draw[orange!85!black, very thick, dashed]
    (0,0)--(1,1)--(2,2)--(3,3)--(4,2)--(5,3)--(6,2)--(7,1)--(8,2)--(9,1)--(10,0);
\node[right, font=\normalsize, orange!85!black] at (5.5, 2.9) {$V^+_p$};
\end{tikzpicture}
\caption{Three representative walks of length $N=10$ starting at $2p-N=0$.
The restricted fortuity window $|2p-N|\leq q-1$ is shaded orange.
The black walk remains confined to the strip and corresponds to an ordinary BPS
state in $\mathcal{B}^p_{\mathrm{o}}$.  The blue dashed walk first exits through the
bottom boundary and contributes to $V^-_p$.  The orange dashed walk
first exits through the top boundary and contributes to $V^+_p$.}
\label{fig:walks}
\end{figure}

In this appendix, we show that the dimensions of the three
sectors of the Hodge
decomposition are each counted by a mutually exclusive class of $\pm1$ walks of length $N$.  The first sector is the ordinary BPS subspace $\mathcal{B}^p_{\mathrm{o}}$, whose dimension equals the number of walks that remain confined within the
restricted fortuity strip for their entire length.  The second is the space $V_p^-$ of non-BPS states annihilated by $Q^\dagger$ but not $Q$, together with the $\mathcal{O}(1)$ exceptional BPS states at level $-q$.  Its dimension equals the number of walks
whose first exit from the strip is through the bottom boundary at $-q$.  The
third is the space $V_p^+$ of non-BPS states annihilated by $Q$ but not $Q^\dagger$,
together with the exceptional BPS states at level $+q$.  The dimension of this space
equals the number of walks first exiting through the top boundary at $+q$. These
three walk classes are illustrated in Figure~\ref{fig:walks}.

\subsection{Preliminaries}
\label{app:prelim}

Fix odd $q\ge3$ and $N\ge1$.  The Hilbert space decomposes as
$\cH=\bigoplus_{p=0}^{N}\cH^p$ with $\dim\cH^p=D_p\equiv\binom{N}{p}$.
The $q$-body supercharge $Q\colon\cH^p\to\cH^{p+q}$ satisfies $Q^2=0$.
We write $\ell\equiv 2p-N$ throughout.

The Hodge decomposition reads
\begin{equation}\label{eq:app-hodge}
  \cH^p \;=\; \mathcal{B}^p \;\oplus\; \operatorname{im}Q\big|_{\cH^{p-q}}
              \;\oplus\; \operatorname{im}Q^\dagger\big|_{\cH^{p+q}},
\end{equation}
where $\mathcal{B}^p = \ker Q|_{\cH^p}\cap\ker Q^\dagger|_{\cH^p}$ is the full BPS
subspace.  We split $\mathcal{B}^p$ into ordinary and exceptional parts,
\begin{equation}
  \mathcal{B}^p \;=\; \mathcal{B}^p_{\mathrm{o}} \;\oplus\; \mathcal{B}^p_{\mathrm{e}},
\end{equation}
where $\mathcal{B}^p_{\mathrm{o}}$ contains the generic BPS states and
$\mathcal{B}^p_{\mathrm{e}}$ the $\mathcal{O}(1)$ states at the boundaries of the fortuity window, $\ell=\pm q$ (Section~\ref{sec:exceptional}). 
The exceptional
piece further splits by boundary,
\begin{equation}
  \mathcal{B}^p_{\mathrm{e}} \;=\; \mathcal{B}^p_{\mathrm{e},-} \;\oplus\; \mathcal{B}^p_{\mathrm{e},+},
\end{equation}
where $\mathcal{B}^p_{\mathrm{e},-}$ contains states only when $\ell=-q$ and
$\mathcal{B}^p_{\mathrm{e},+}$ an equal number of states when $\ell=+q$.

We absorb each exceptional piece into a non-BPS sector, defining
\begin{align}
  V_p^- &\;:=\; \operatorname{im}Q^\dagger\big|_{\cH^{p+q}} \;\oplus\; \mathcal{B}^p_{\mathrm{e},-}, \label{eq:Vminus-def}\\
  V_p^+ &\;:=\; \operatorname{im}Q\big|_{\cH^{p-q}} \;\oplus\; \mathcal{B}^p_{\mathrm{e},+}. \label{eq:Vplus-def}
\end{align}
The non-BPS states in $V_p^-$ are annihilated by $Q^\dagger$ but not $Q$, and those in $V_p^+$ are annihilated by $Q$ but not $Q^\dagger$. The exceptional states, being BPS, are annihilated by both. With this absorption, the decomposition~\eqref{eq:app-hodge} becomes
\begin{equation}\label{eq:hodge_final}
  \cH^p \;=\; \mathcal{B}^p_{\mathrm{o}} \;\oplus\; V_p^- \;\oplus\; V_p^+,
\end{equation}
with dimension count
\begin{equation}\label{eq:dimcount}
  h^p_{\mathrm{o}} \;\equiv\; \dim\mathcal{B}^p_{\mathrm{o}}
  \;=\; D_p - \dim V_p^- - \dim V_p^+.
\end{equation}

The primary analytic tool throughout is the index-truncated dimension, originally introduced in \cite{CCSY}. It is defined for $p\in\mathbb{Z}$ by
\begin{equation}\label{eq:Dtil}
  \Dtil_p \;:=\; \sum_{n\ge 0}(-1)^n\binom{N}{p-nq},
\end{equation}
with $\binom{N}{k}=0$ for $k<0$ or $k>N$, so $\Dtil_p=0$ for $p<0$.
Its generating function, as a formal power
series, is \cite{Wilf}
\begin{equation}\label{eq:Dtil_gf}
  \sum_{p\ge0}\Dtil_p\,x^p \;=\; \frac{(1+x)^N}{1+x^q}.
\end{equation}
To see this, substitute the definition~\eqref{eq:Dtil} and set $j=p-nq$:
\begin{equation}
  \sum_{p}\Dtil_p\,x^p = \sum_{n\ge 0}(-1)^n\sum_{j=0}^{N}\binom{N}{j}\,x^{j+nq} = \sum_{n\ge 0}(-x^q)^n\sum_{j=0}^{N}\binom{N}{j}\,x^j.
\end{equation}
The inner sum is $(1+x)^N$ by the binomial theorem and the outer sum is
$(1+x^q)^{-1}$ as a formal geometric series, giving~\eqref{eq:Dtil_gf}.
Multiplying both sides by $1+x^q$ and extracting the coefficient of $x^p$ gives
the \emph{splitting identity}
\begin{equation}\label{eq:split}
  D_p \;=\; \Dtil_p + \Dtil_{p-q}.
\end{equation}

Let $M_p$ be the matrix representation of $Q\colon\cH^p\to\cH^{p+q}$. With this identification, the dimensions of $V_p^-$ and $V_p^+$ are
\begin{align}
  \dim V_p^- &= \rk M_p + \dim\mathcal{B}^p_{\mathrm{e},-}, \\
  \dim V_p^+ &= \rk M_{p-q} + \dim\mathcal{B}^p_{\mathrm{e},+}. \label{eq:dim_Vp_plus_rank}
\end{align}
In \cite{CCSY}, it was conjectured that the rank of $M_p$ is given by $\Dtil_p$ for $\ell<1-q$. In this appendix, we extend this conjecture to the rank of $M_p$ within and above the fortuity window, including corrections coming from exceptional states. This rank formula is the key input assumption of this appendix, provided below.

\begin{conjecture}[Rank formula\footnote{This conjectural formula is closely related to questions studied in commutative algebra. The supercharge acts by multiplication with an odd form in an exterior algebra, and the ranks of its charge-sector blocks are consequently related to the Hilbert series of the corresponding principal quotient. Generic principal ideals of this type have been studied in~\cite{MorenoSociasSnellman,LundqvistNicklasson}, but their Hilbert series are not known in general. It would be interesting to clarify the precise relation between these results and the walk description developed here.}]\label{conj:rank}
For generic couplings, the rank of $M_p$ is given by
\begin{equation}\label{eq:rank-formula}
  \dim V_p^- \;=\; \rk M_p + \dim \mathcal{B}^p_{\mathrm{e},-} \;=\;
  \begin{cases}
    \Dtil_p,       & \ell < 1-q,     \\[4pt]
    \Dtil_{N-p-q}, & \ell \geq 1-q.
  \end{cases}
\end{equation}
That is, the rank of $M_p$ determines the dimension of $V_p^-$, up to corrections owing to exceptional states. This formula has been verified by exact diagonalization using multiple realizations of the random couplings at odd $q$ from $3$ to $9$ and from $N=q$ up to $14$.
\end{conjecture}

The dimension of $V_p^+$ then follows from the rank formula. Starting from~\eqref{eq:dim_Vp_plus_rank}:
\begin{align}
    \dim V_p^+ &= \rk M_{p-q} + \dim\mathcal{B}^p_{\mathrm{e},+} \\
    &= \rk M_{p-q} + \dim\mathcal{B}^{p-q}_{\mathrm{e},-} \\
    &= \dim V_{p-q}^-. \label{eq:dim_Vp_plus_final}
\end{align}
The second line used the fact that the shift $p\mapsto p-q$ lowers the level
$\ell = 2p-N$ by $2q$, carrying the upper window boundary $\ell=+q$ to the
lower one $\ell=-q$. Another way of obtaining the dimension of $V_p^+$ is by substituting $p\mapsto N-p$ into the rank formula, a symmetry we call \emph{palindromic symmetry}:
\begin{equation}\label{eq:Vplus_dim}
  \dim V_p^+ \;=\; \dim V_{N-p}^-. 
\end{equation}
In other words, the rank formula for the $Q$-annihilated sector is
not independent. Instead, it is entirely determined by the rank formula for the $Q^\dagger$-annihilated sector via $p\mapsto N-p$.

Plugging~\eqref{eq:rank-formula} and \eqref{eq:dim_Vp_plus_final} into the dimension count~\eqref{eq:dimcount}, and using the splitting identity, the ordinary BPS count becomes
\begin{equation}
  h^p_{\mathrm{o}} = D_p - \Dtil_{p-q} - \Dtil_{N-p-q} = \Dtil_p - \Dtil_{N-p-q}. \label{eq:h^p_Dtil_split}
\end{equation}
This derivation used the fact that when $p$ is inside the restricted fortuity
window, $p-q$ lies strictly below it so $\dim V_{p}^+=\Dtil_{p-q}$. The result of Appendix~\ref{app:strip} will be to show that $S_p$,
the number of strip-confined walks, equals precisely $\Dtil_p-\Dtil_{N-p-q}$,
establishing $S_p=h^p_{\mathrm{o}}$ conditionally on Conjecture~\ref{conj:rank}.

We next establish the walk encoding and its combinatorial symmetry.
A basis state $(n_1,\ldots,n_N)\in\{0,1\}^N$ encodes as a walk
$(s_1,\ldots,s_N)$ with $s_i=2n_i-1\in\{\pm1\}$ and position
$x_k=\sum_{i=1}^k s_i$, $x_0=0$.  The fermion number is $p=\#\{i:s_i=+1\}$
and the endpoint is $x_N=\ell=2p-N$.
The strip is $\W=\{x\in\mathbb{Z}:|x|\leq q-1\}$ and the
first-exit time is $\tau(w)=\min\{k\geq1:|x_k|\geq q\}$,
with $\tau=\infty$ if the walk never leaves $\W$.
The three walk classes are
\begin{align*}
  S_p &:= \#\{w\in\cH^p:\tau=\infty\}, \\
  B_p &:= \#\{w\in\cH^p:x_\tau=-q\},   \\
  T_p &:= \#\{w\in\cH^p:x_\tau=+q\},
\end{align*}
and every walk falls in exactly one class, so $D_p = S_p + B_p + T_p$.

The walks exhibit a palindromic symmetry mirroring the state-space symmetry above.
Define the involution $\iota(s_1,\ldots,s_N):=(-s_1,\ldots,-s_N)$, an instance of
the reflection principle \cite{Feller}. We show $T_p=B_{N-p}$; the proof of $S_p=S_{N-p}$ is identical.
Let $w=(s_1,\ldots,s_N)\in\cH^p$ and $\bar{w}=\iota(w)$.
The fermion number of $\bar{w}$ is $\bar{p}=\#\{i:-s_i=+1\}=N-p$, so
$\bar{w}\in\cH^{N-p}$.  The positions satisfy $\bar{x}_k=-x_k$ for all $k$,
so $|\bar{x}_k|=|x_k|$ and the first-exit time is preserved: $\bar{\tau}=\tau$.
Since $\bar{x}_{\bar\tau}=-x_\tau$, a top exit becomes a bottom exit.
Therefore $\iota$ maps each $w\in\cH^p$ with $x_\tau=+q$ bijectively to a
walk $\bar{w}\in\cH^{N-p}$ with $\bar{x}_{\bar\tau}=-q$, giving $T_p=B_{N-p}$.

\subsection{Ordinary BPS states are strip-confined walks}
\label{app:strip}

\providecommand{\Z}{\mathbb{Z}}

We now show directly that the ordinary BPS count
$h^p_{\mathrm{o}}=\Dtil_p-\Dtil_{N-p-q}$ of~\eqref{eq:h^p_Dtil_split} equals
the number of strip-confined
walks, by specializing the two-boundary enumeration of
Krattenthaler~\cite{Krattenthaler}. Theorem~10.3.3 of~\cite{Krattenthaler}
counts monotone lattice paths from $(a,b)$ to $(c,d)$, with unit east and
north steps, staying weakly between the lines $y=x+s$ and $y=x+t$. For
$a+s\le b\le a+t$ and $c+s\le d\le c+t$, the number $L$ of such paths is
\begin{equation}
  L=\sum_{k\in\Z}\left[
      \binom{c+d-a-b}{\,c-a-k(t-s+2)\,}
     -\binom{c+d-a-b}{\,c-b-k(t-s+2)+t+1\,}
    \right].
  \label{eq:krat1033}
\end{equation}
Under the $45^\circ$ rotation sending up-steps to north steps and down-steps
to east steps, a $\pm1$ walk of length $N$ from height $0$ to $\ell=2p-N$
confined to $\lvert x\rvert\le q-1$ becomes such a path with endpoint
$(N-p,\,p)$. Setting accordingly $(a,b)=(0,0)$, $(c,d)=(N-p,p)$, $t=q-1$,
$s=1-q$, Equation~\eqref{eq:krat1033} becomes
\begin{equation}
  L=\sum_{k\in\Z}\left[
      \binom{N}{N-p-2kq}-\binom{N}{N-p-2kq+q}
    \right].
  \label{eq:krat-sub}
\end{equation}
We split this sum at $k=1$ into $L=L_-+L_+$, where $L_-$ collects the terms
with $k\le0$ and $L_+$ those with $k\ge1$.

In $L_-$, write $k=-j$ with $j\ge0$ and apply $\binom{N}{i}=\binom{N}{N-i}$ to
both binomials. Splitting the result by the parity of $j$,
\begin{align}
  L_-&=\sum_{j\ge0}\left[\binom{N}{p-2jq}-\binom{N}{p-(2j+1)q}\right]
  \nonumber\\
  &=\sum_{\substack{n\ge0\\ n\ \mathrm{even}}}(-1)^n\binom{N}{p-nq}
   +\sum_{\substack{n\ge0\\ n\ \mathrm{odd}}}(-1)^n\binom{N}{p-nq}
   =\Dtil_p .
  \label{eq:half-Dp}
\end{align}
In $L_+$, split by parity:
\begin{align}
  L_+&=\sum_{k\ge1}\left[\binom{N}{N-p-2kq}-\binom{N}{N-p-(2k-1)q}\right]
  \nonumber\\
  &=-\sum_{\substack{n\ge0\\ n\ \mathrm{odd}}}(-1)^n\binom{N}{N-p-(n+1)q}
    -\sum_{\substack{n\ge0\\ n\ \mathrm{even}}}(-1)^n\binom{N}{N-p-(n+1)q}
  \nonumber\\
  &=-\sum_{n\ge0}(-1)^n\binom{N}{N-p-q-nq}=-\Dtil_{N-p-q}.
  \label{eq:half-Dnpq}
\end{align}
Adding the two halves gives $L=\Dtil_p-\Dtil_{N-p-q}=h^p_{\mathrm{o}}$,
so $h^p_{\mathrm{o}}$ is directly the number of $\pm1$ walks of length $N$
from $0$ to $\ell=2p-N$ confined to $\lvert x\rvert\le q-1$:
\begin{equation}
  S_p \;=\; \Dtil_p-\Dtil_{N-p-q} \;=\; h^p_{\mathrm{o}}\,,
\end{equation}
the first equality unconditional, the second conditional on
Conjecture~\ref{conj:rank} through~\eqref{eq:h^p_Dtil_split}.

\subsection{\texorpdfstring{$V_p^\pm$}{V\_p\^{}+/-} states are first-exit walks}
\label{app:firstexit}

We now show that $B_p=\dim V_p^-$ and $T_p=\dim V_p^+$: the $Q^\dagger$-annihilated
sector is counted by walks first exiting through $-q$, and the
$Q$-annihilated sector by walks first exiting through $+q$, in each
case including the relevant exceptional BPS states.

A walk in $B_p$ decomposes into a prefix of length $\tau$ staying in the strip $\W$ and
first exiting through $-q$, followed by a suffix of $N-\tau$ unconstrained steps
from $-q$ to $\ell$ \cite{BanderierFlajolet,BanderierWallner,Ouvry}.  Since prefix and suffix are independent,
\begin{equation}
  B_p \;=\; \sum_{\tau} f_\tau \binom{N-\tau}{\tfrac{N-\tau+\ell+q}{2}},
  \label{eq:Bp_conv}
\end{equation}
where $f_\tau$ is the number of walks of length $\tau$ from $0$ first reaching
$-q$ in $\W$, and the binomial counts the number of free walks of $N-\tau$ steps
from $-q$ to $\ell$: in $N-\tau$ steps, one needs $(N-\tau+\ell+q)/2$ up-steps to
achieve a net displacement of $\ell+q$.
Since $x_0=0$ is even and $-q$ is odd, $\tau$ must be odd; writing $\tau=q+2j$
and substituting:
\begin{equation}
  B_p \;=\; \sum_{j=0}^{M}
  f_{q+2j}\binom{N-q-2j}{p-j},
  \qquad M:=\left\lfloor\frac{N-q}{2}\right\rfloor.
  \label{eq:Bp_j}
\end{equation}
The sum terminates at $j=M$ because $\tau=q+2j\leq N$ requires $j\leq(N-q)/2$,
i.e.\ $j\leq M$.

Let $A_\W$ be the adjacency matrix of the path graph $P_{2q-1}$ on $\W$. The only vertex of
$\W$ adjacent to $-q$ is $-(q-1)$, so any prefix walk ending at $-q$ must have
$x_{\tau-1}=-(q-1)$. The walk therefore takes $q+2j-1$ steps from $0$ to $-(q-1)$
staying entirely within $\W$, followed by one forced step to $-q$. Because the starting point $x=0$ corresponds
to vertex $q$ in the path graph, the walk count is given by the matrix element $(A_{\W}^{q+2j-1})_{kl}$ with $k=q$ and $l=1$, namely
\begin{equation}
  f_{q+2j} \;=\; \bigl(A_\W^{q+2j-1}\bigr)_{q,1}.
  \label{eq:ftau_matrix}
\end{equation}
The spectrum of adjacency matrices is well-known \cite{FelsnerHeldt,Biggs}. For a path graph of length $m$, the eigenvalues are $\mu_k = 2\cos(k\pi/(m+1))$, $k=1,\ldots,m$. 

Assemble the matrix elements into the per-step walk generating function
\begin{equation}
  G(z) \;:=\; \sum_{n\ge 0} \bigl(A_\W^{\,n}\bigr)_{q,1}\, z^{n}
        \;=\; \bigl[(I-zA_\W)^{-1}\bigr]_{q,1}.
\end{equation}
The walk must end at height $-(q-1)$, which is even for odd $q$. Because the graph is bipartite, $\bigl(A_\W^{\,n}\bigr)_{q,1}=0$ for odd $n$. Moreover, $\bigl(A_\W^{\,n}\bigr)_{q,1}=0$ for $n<q-1$. Therefore, $G(z)$ can be written
\begin{equation}
  G(z) = \sum_{j\ge 0}\bigl(A_\W^{\,q-1+2j}\bigr)_{q,1}\,z^{q-1+2j}
       = z^{\,q-1}\sum_{j\ge 0} f_{q+2j}\,z^{2j}
       = z^{\,q-1}\,F(z^{2}),
\end{equation}
introducing $F(t):=\sum_{j\geq0}f_{q+2j}t^j$. With $t=z^{2}$, $F(t)$ is given by the matrix resolvent
\begin{equation}
  F(t) \;=\; z^{-(q-1)}\,\bigl[(I-zA_\W)^{-1}\bigr]_{q,1}.
  \label{eq:F-resolvent}
\end{equation}

The matrix $M:=I-zA_\W$ is a $(2q-1)\times(2q-1)$ symmetric tridiagonal matrix with $1$ on the diagonal and $-z$ on the two off-diagonals. Using the Usmani tridiagonal-inverse formula, the inverse element is \cite{Usmani1994}
\begin{equation}
  \bigl(M^{-1}\bigr)_{q,1}
   = (-1)^{q+1}\Bigl(\textstyle\prod_{i=1}^{q-1} M_{i,i+1}\Bigr)\,
     \frac{\theta_{0}\,\phi_{q+1}}{\det M},
\end{equation}
with $\theta_{0}=1$ and $\phi_{q+1}=\det M[\,q{+}1,\dots,2q{-}1\,]$ the determinant of the trailing $(q-1)\times(q-1)$ block. Since $M_{i,i+1}=-z$ and $q$ is odd, the prefactor becomes $z^{\,q-1}$. Let $U_{m}(t)$ represent the determinant of the size-$m$ block, with $1$ on the diagonal and $-z$ on the off-diagonals. Then $\det M=U_{2q-1}(t)$ and $\phi_{q+1}=U_{q-1}(t)$. Substituting into
\eqref{eq:F-resolvent}, the $z^{\,q-1}$ factors cancel and
\begin{equation}
  F(t) \;=\; \frac{U_{q-1}(t)}{U_{2q-1}(t)}\,.
  \label{eq:F-continuant}
\end{equation}

Define the denominator polynomial as $D(t):=1/F(t)$. The determinant vanishes when $1/z$ is an eigenvalue of the path-graph adjacency matrix. Pairing the eigenvalues $\mu_k$ and $-\mu_k$ and writing $t=z^2$, the distinct finite roots are $t=1/\mu_k^2$ for $k=1,\ldots,\lfloor m/2\rfloor$. Since $U_m(0)=1$, this gives
\begin{equation}
  U_{m}(t)=\prod_{k=1}^{\lfloor m/2\rfloor}\Bigl(1-4\cos^{2}\left(\tfrac{k\pi}{m+1}\right)\,t\Bigr).
\end{equation}
For $m=2q-1$, the roots are $\lambda_{k}^{2}=\left(2\cos(k\pi/2q)\right)^2$, $k=1,\dots,q-1$. For
$m=q-1$, they are $\lambda_{2k}^{2}$, $k=1,\dots,(q-1)/2$, which is precisely the even-index subset. Dividing $U_{2q-1}$ by $U_{q-1}$, only the odd-index factors $K=\{1,3,\dots,q-2\}$ are retained:
\begin{equation}
  D(t)=\frac{U_{2q-1}(t)}{U_{q-1}(t)}=\prod_{k\in K}\bigl(1-\lambda_{k}^{2}\,t\bigr).
  \label{eq:D_pf}
\end{equation}

We claim that $D$ satisfies the identity $D\bigl(x/(1+x)^2\bigr)=(1+x^q)/(1+x)^q$,
equivalently
\begin{equation}\label{eq:F_pf_final}
  (1+x)^{N-q}F\!\left(\frac{x}{(1+x)^2}\right) \;=\; \frac{(1+x)^N}{1+x^q},
\end{equation}
which follows from $F(t)=1/D(t)$.
To prove the identity for $D$, substitute~\eqref{eq:D_pf} to get
\begin{equation}
  D\!\left(\frac{x}{(1+x)^2}\right)
  \;=\; \prod_{k\in K}\!\left(1-\frac{x\lambda_k^2}{(1+x)^2}\right)
  \;=\; \frac{\mathcal{N}(x)}{(1+x)^{q-1}},
\end{equation}
where $\mathcal{N}(x)$ is defined as
\begin{equation}\label{eq:N_def}
  \mathcal{N}(x) \;:=\; \prod_{k\in K}\bigl(x^2+(2-\lambda_k^2)x+1\bigr).
\end{equation}
Each factor in~\eqref{eq:N_def} is monic of degree~$2$, so $\mathcal{N}$ has degree $q-1$
with leading coefficient~$1$.  At $x=-1$ each factor equals $\lambda_k^2>0$,
confirming the denominator $(1+x)^{q-1}$ does not cancel.  The $k$-th factor
vanishes at a root $\rho_k$ of $1+x^q$, and also at
$1/\rho_k=\rho_{2q-k}$. Therefore, $\mathcal{N}$ vanishes at all $q-1$ roots $\rho_k\neq-1$.

Since $\mathcal{N}$ has degree $q-1$ and $q-1$ prescribed zeros, we can write
$\mathcal{N}(x)=c\prod_k(x-\rho_k)$ for some constant $c$.  The $\rho_k$ are exactly
the roots of $1-x+\cdots+x^{q-1}=(1+x^q)/(1+x)$, so the factor theorem gives
$\prod_k(x-\rho_k)=1-x+\cdots+x^{q-1}$.  Evaluating at $x=0$ via~\eqref{eq:N_def}
gives $\mathcal{N}(0)=\prod_{k\in K}1=1$, which fixes $c=1$.  Therefore
\begin{equation}
  D\!\left(\frac{x}{(1+x)^2}\right)
  \;=\; \frac{1-x+\cdots+x^{q-1}}{(1+x)^{q-1}}
  \;=\; \frac{1+x^q}{(1+x)^q}.
\end{equation}

With $F_N(t)=\sum_{j=0}^{M}f_{q+2j}t^j$ denoting the truncated generating
function, the generating function of $B_p$ is
\begin{equation}
  \sum_{p=0}^N B_p\,x^p
  \;=\; (1+x)^{N-q}\cdot F_N\!\left(\frac{x}{(1+x)^2}\right)
  \;=\; \sum_{j=0}^{M} f_{q+2j}\,x^j(1+x)^{N-q-2j}.
  \label{eq:Bp_gf}
\end{equation}
To derive this, set $r=p-j$ in~\eqref{eq:Bp_j} to reindex the sum over $p$:
\begin{align}
  \sum_p B_p x^p
  &\;=\; \sum_{j=0}^M f_{q+2j}\sum_p\binom{N-q-2j}{p-j}x^p \nonumber\\
  &\;=\; \sum_{j=0}^M f_{q+2j}\,x^j\sum_{r=0}^{N-q-2j}\binom{N-q-2j}{r}x^r \nonumber\\
  &\;=\; \sum_{j=0}^M f_{q+2j}\,x^j(1+x)^{N-q-2j},
\end{align}
where the last step is the binomial theorem, valid since $N-q-2j\geq0$ for
$j\leq M$.  Writing $F_N=F-(F-F_N)$ and applying~\eqref{eq:F_pf_final}:
\begin{equation}\label{eq:split_gf}
  \sum_p B_p x^p
  \;=\; \frac{(1+x)^N}{1+x^q}
  \;-\; \underbrace{\sum_{j>M} f_{q+2j}\,x^j(1+x)^{N-q-2j}}_{=:\,T(x)},
\end{equation}
where $T(x)$ is a rational function with poles at $x=-1$.

We show $B_p=\dim V_p^-$ by comparing $[x^p]$ on both sides
of~\eqref{eq:split_gf}, using $[x^p]\,(1+x)^N/(1+x^q)=\Dtil_p$ and the sector
dimension~\eqref{eq:rank-formula}.

\medskip
\textbf{Range 1: $\ell<1-q$, equivalently $p\leq M$.}
Each term of $T(x)$ has index $j>M$, so $p-j\leq M-(M+1)=-1<0$.
Hence $[x^p]\,x^j(1+x)^{N-q-2j}=\binom{N-q-2j}{p-j}=0$, giving $[x^p]T(x)=0$.
Therefore $B_p=[x^p](1+x)^N/(1+x^q)=\Dtil_p=\dim V^-_p$.

\medskip
\textbf{Range 2: $\ell\geq1-q$, equivalently $p>M$.}
We split into two sub-cases.

\medskip
\emph{Sub-case 2a: $M<p\leq N-q$.}
Set $p':=N-q-p$.  Since $p\geq M+1$ and $M=\lfloor(N-q)/2\rfloor$, we have
$p'\leq N-q-M-1\leq M$, and $p\leq N-q$ gives $p'\geq0$.
For each $j=0,\ldots,M$, the quantity $N-q-2j\geq0$ since $j\leq M$ implies
$2j\leq N-q$.  Applying $\binom{n}{k}=\binom{n}{n-k}$ with $k=p-j$ and
$n-k=p'-j$:
\begin{equation}
  B_p \;=\; \sum_{j=0}^M f_{q+2j}\binom{N-q-2j}{p-j}
       \;=\; \sum_{j=0}^M f_{q+2j}\binom{N-q-2j}{p'-j}
       \;=\; B_{p'}.
\end{equation}
Since $p'\leq M$, Range~1 gives $B_{p'}=\Dtil_{p'}=\Dtil_{N-p-q}=\dim V^-_p$.

\medskip
\emph{Sub-case 2b: $p>N-q$.}
For each $j=0,\ldots,M$: since $p>N-q\geq N-q-j$ and $j\geq0$, we have
$p-j>N-q-2j$, so $\binom{N-q-2j}{p-j}=0$, and therefore $B_p=0$.  When
$p>N-q$ we have $N-p-q<0$, so all terms in
$\Dtil_{N-p-q}=\sum_{n\geq0}(-1)^n\binom{N}{N-p-q-nq}$ vanish, giving
$B_p=0=\dim V^-_p$.

\medskip
This completes the proof that $B_p=\dim V_p^-$, conditional on
Conjecture~\ref{conj:rank}. The next step is to show that $T_p=\dim V_p^+$.
One way to see this is from the partition $D_p=B_p+T_p+S_p$: since $S_p$
corresponds to the ordinary BPS states and $B_p$ to $\dim V_p^-$, the remaining
$T_p$ must equal $\dim V_p^+$.  However, this argument relies on knowledge of
$S_p$, proved in Appendix~\ref{app:strip}.  A more direct route uses only $B_p$ and palindromic
symmetry.  By palindromic symmetry of walks, $T_p=B_{N-p}$.
By the above, $B_{N-p}=\dim V_{N-p}^-$.
By palindromic symmetry on the level of states, $\dim V_{N-p}^-=\dim V_p^+$.
Therefore $T_p=\dim V_p^+$.

\begin{figure}[tb]
\begin{center}
\includegraphics[width=7.6cm]{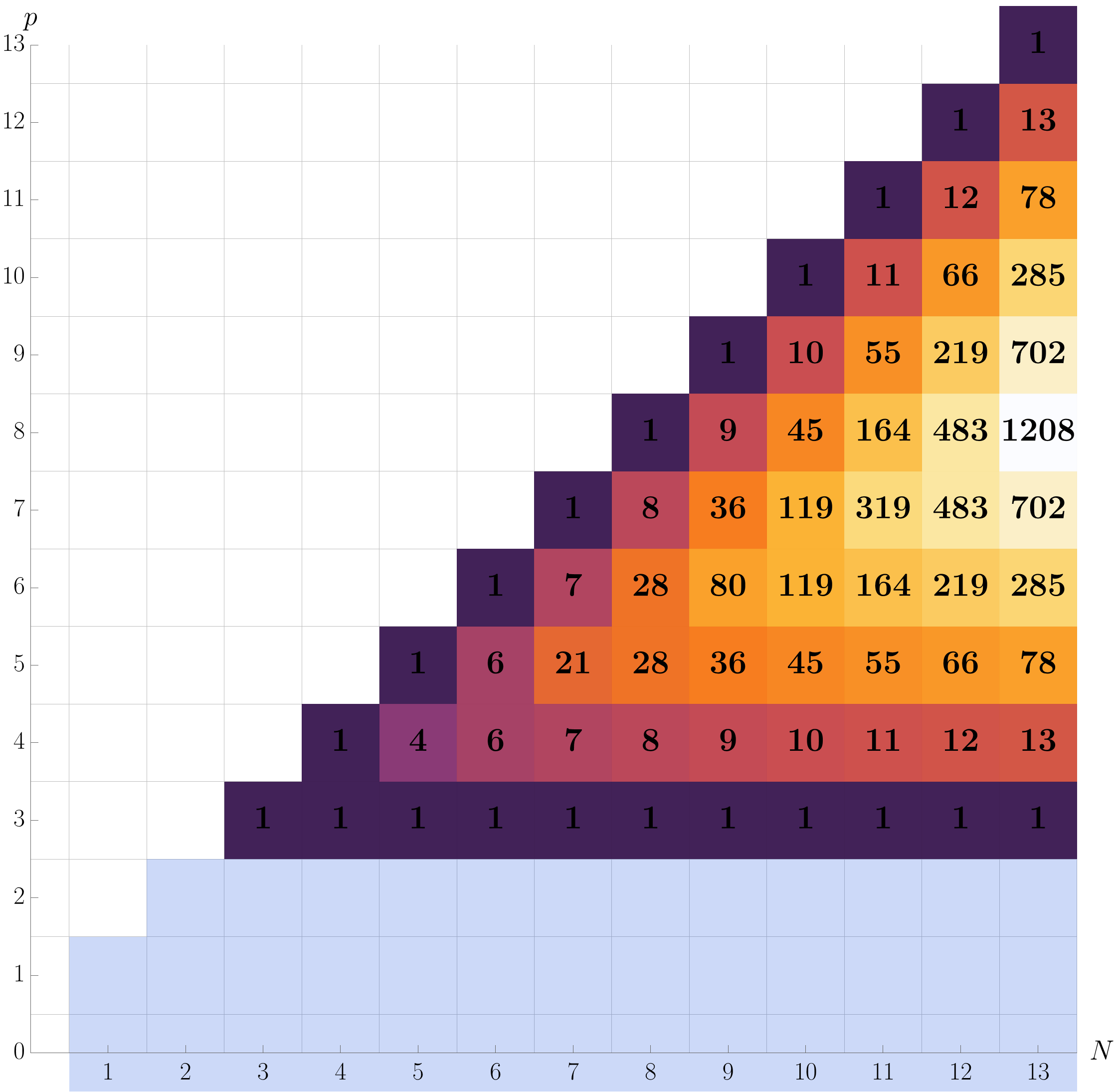} \ \ \
\includegraphics[width=7.6cm]{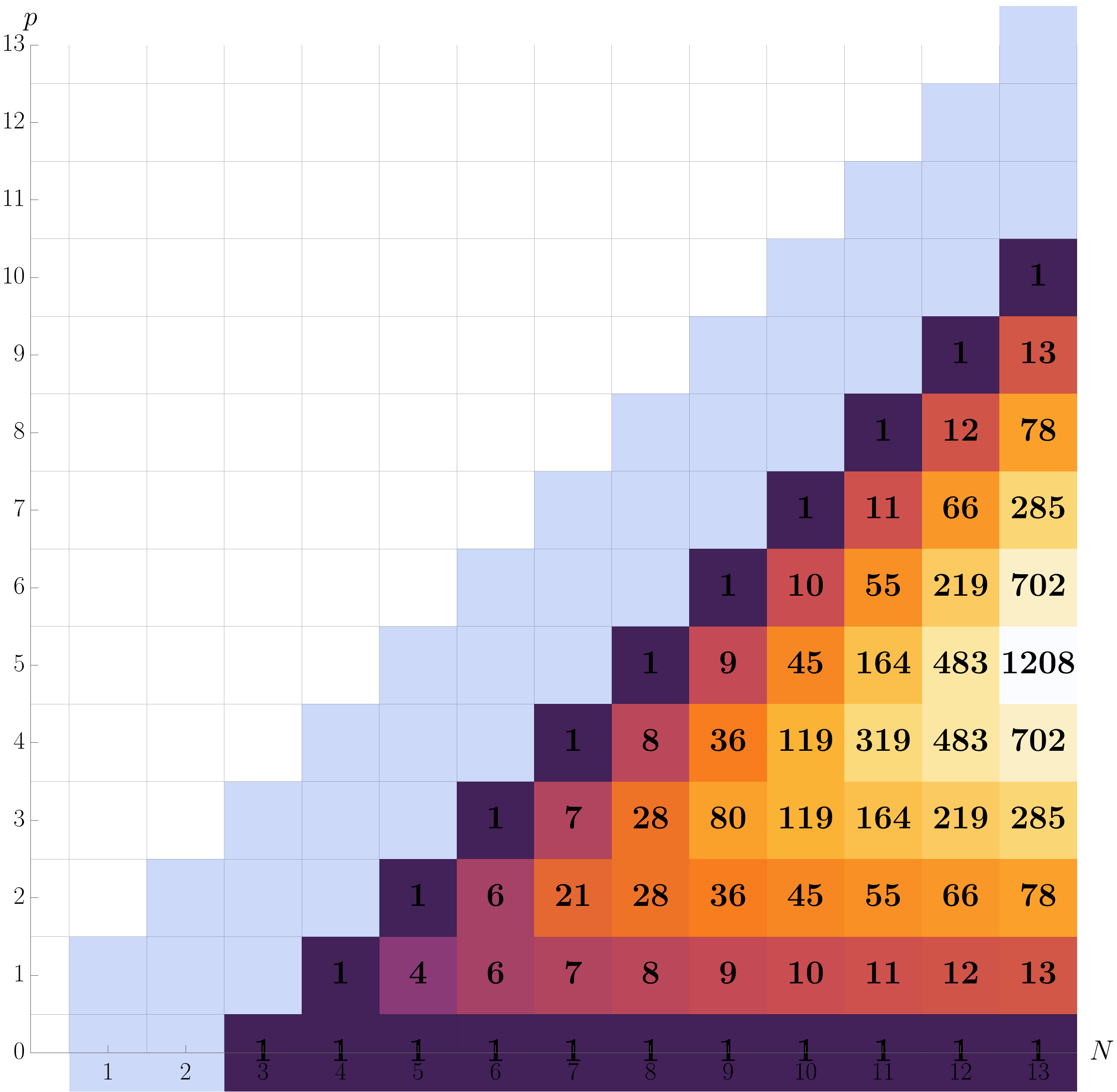}
\caption{\label{fig:nonbpsdata}
{\it Left:} Number of non-BPS states annihilated by $Q$ but not by
$Q^\dagger$: the $\operatorname{im}Q$ summand of $V^+_p$, counted away
from the exceptional boundaries by top-exit walks $T_p$.
{\it Right:} Number of non-BPS states annihilated by $Q^\dagger$ but not
by $Q$: the $\operatorname{im}Q^\dagger$ summand of $V^-_p$, counted by
bottom-exit walks $B_p$. The numbers match those in Appendix~B
of~\cite{KanazawaWettig}.}
\end{center}
\end{figure}

More explicitly, one can show $T_p=B_{p-q}$ directly.
For $p<q$: $T_p=0$ since the walk has only $p<q$ up-steps and cannot reach $+q$,
and $B_{p-q}=0$ since $p-q<0$.  For $q\leq p\leq N$, setting $n:=N-q-2j$ and
$k:=(p-q)-j$, one checks $n-k=(N-p)-j$, so by $\binom{n}{n-k}=\binom{n}{k}$:
\begin{equation}
  B_{N-p} \;=\; \sum_{j=0}^M f_{q+2j}\binom{N-q-2j}{(N-p)-j}
           \;=\; \sum_{j=0}^M f_{q+2j}\binom{N-q-2j}{(p-q)-j}
           \;=\; B_{p-q}.
\end{equation}
Hence $T_p=B_{N-p}=B_{p-q}=\dim V_p^+$.

To summarize, under the walk encoding the decomposition~\eqref{eq:hodge_final}
corresponds precisely to the partition $D_p=S_p+B_p+T_p$ of all $\pm1$ walks of
length $N$ into strip-confined, first-exit-bottom, and first-exit-top
walks---unconditionally for $B_p$ below the window, and conditionally on
Conjecture~\ref{conj:rank} for the remaining identifications.
The dimension of $V_p^-$ counts states in $\operatorname{im}Q^\dagger$ together with the
$\ell=-q$ exceptional BPS states, and is enumerated by bottom-exit walks;
by palindromic symmetry on the
level of states, $\dim V_p^+=\dim V_{N-p}^-$, so the $Q$-annihilated
sector $V_p^+$ together with its $\ell=+q$ exceptional states is counted by walks
first exiting through $+q$.\qed

\medskip
The identifications are confirmed by exact diagonalization.
Figure~\ref{fig:nonbpsdata} displays the measured dimensions of the two
non-BPS summands for $q=3$ and generic couplings. Away from the boundaries
$|\ell|=q$ they coincide with the first-exit counts: $B_p$ for the
$\operatorname{im}Q^\dagger$ summand of $V^-_p$ and $T_p$ for the
$\operatorname{im}Q$ summand of $V^+_p$. At $|m|=q$ the walk counts exceed
the non-BPS dimensions by exactly the $\mathcal{O}(1)$ exceptional
multiplicities, as required by the absorption
in~\eqref{eq:Vminus-def}--\eqref{eq:Vplus-def}. The same dimensions appear
in the random-matrix classification of Kanazawa and
Wettig~\cite{KanazawaWettig}.

\clearpage

\bibliographystyle{JHEP}
\bibliography{paper}

\end{document}